# A Model of the Globally-averaged Thermospheric Energy Balance

Karthik Venkataramani [*, 1], Scott. M. Bailey [1], Srimoyee Samaddar[1], Justin Yonker[1]

[1] Center for Space Science and Engineering, Bradley Department of Electrical and Computer Engineering, Blacksburg, Virginia, USA.

Corresponding Author: Karthik Venkataramani ([karthik0@vt.edu](mailto:karthik0@vt.edu))

*This work was done while the author was at Virginia Tech

Code associated with this work can be found at https://github.com/kvenkman/ace1d

## Key Points

1. We develop a first principles based, self-consistent, 1D model that calculates the mean state of the Earth's thermosphere and ionosphere

2. Using an up-to-date chemistry scheme and specifications of various processes, the model allows for detailed energy balance calculations.

3. Modeled exospheric temperatures are, on average, within 10% of MSIS values, and nitric oxide densities are within 25% of measurements.

# Abstract

The Atmospheric Chemistry and Energetics (ACE) 1D model is a first principles based model that generates a globally averaged thermosphere and ionosphere in terms of constituent major, minor, and charged species, as well as associated temperatures. The model solves the 1D continuity and energy equations representing relevant physical processes, and is supported by a chemistry scheme that reflects our current understanding of chemical processes in the region. The model is a first in its detailed treatment of nitric oxide (NO) chemistry, including the $N_2(A)$ + O reaction as a source, and accounting for chemiluminescence effects resulting from the vibrationally excited NO produced by $N(^2D, ^4S)$ + $O_2$. The model utilizes globally averaged solar fluxes between 0.05-175 nm as the primary form of energy input, parameterized using the F10.7 index to reflect variations over the course of a solar cycle. The model also includes joule heating effects, magnetospheric fluxes, and a parameterized treatment of photoelectron effects as secondary heating sources. The energy inputs are balanced by radiative losses from the neutral thermosphere in the form of infrared emissions from $CO_2$, NO and $O(^3P)$. Atmospheric profiles are generated for a solar cycle, and are compared with empirical model results as well as observational data. On average, calculated exospheric temperatures are within 10% of MSIS values, while peak electron densities are within a factor of 2 of IRI values. The model is shown to reproduce the NO peak at 106 km, and densities within 25% of globally averaged satellite measurements.

# Plain Language Summary

While atmospheric models have tended towards increasing complexity, one-dimensional models remain a crucial tool in the study of Earth's atmosphere. Such models provide a mean representation of the region of study, and serve as a useful testbed for improved numerical descriptions of key processes and new laboratory results. The Atmospheric Chemistry and Energetics (ACE) 1D model is presented in this vein. In particular, the model is a first in its comprehensive and up-to-date treatment of nitric oxide (NO) chemistry and energetics, which plays an important role in determining the composition and energy balance of the region. The ACE1D model generates altitude profiles of the Earth's thermosphere and ionosphere in terms of its constituent species and temperatures in an aggregate sense, and is able to do so for a wide range of inputs that are expected over the course of a solar cycle. The model results are shown to agree well with observations and results from established empirical models, reproducing exospheric temperatures to within 10% of empirically derived values, and peak electron densities within a factor of 2. The model is also shown to reproduce a characteristic NO density peak near 106 km, and values within 25% of available measurements.

## 1. Introduction

Numerical modeling has been a valuable tool that has complemented observations of the Earth's upper atmosphere to improve our understanding of the region and processes occurring therein. The fidelity with which atmospheric models reproduce observational features reflect our understanding of the processes, while also serving to highlight gaps in our knowledge that motivate further study.

For first principles-based models, we begin with the laws of fluid mechanics and thermodynamics expressed as partial differential equations, which are solved to obtain densities and temperatures for the domain under consideration. In comparison, for empirical models, we begin with a set of observations to which we fit approximating functions, optimizing for the goodness of fit. The fitting coefficients in such empirical models are periodically updated to reflect the availability of newer data. Prevalent models of the Earth's upper atmosphere are rooted in early development efforts of 1D and 2D physical models of planetary thermospheres (Dickinson, 1971; Dickinson et al., 1975; Dickinson and Ridley, 1972; Dickinson and Ridley, 1975), which subsequently evolved into efforts of modeling the Earth's upper atmosphere (Dickinson et al., 1981; Dickinson et al., 1984; Roble, 1987; Roble, 1995). This paralleled efforts to develop empirical models of the neutral thermosphere (Hedin, 1983; Picone et al., 2002; Jacchia, 1977) and the ionosphere (Schunk et al., 2004; Rawer et al., 1978).

The availability of increased computational power subsequently led to the development and adoption of three dimensional atmospheric models. In the thermosphere/ionosphere community, these include Thermospheric-Ionosperic-Electrodynamics General Circulation Model (TIE-GCM; Qian et al., 2014), Global Ionosphere Thermosphere Model (GITM; Ridley et al., 2006), SAMI3 (Huba et al., 2008) among others, as well as recent efforts to expand models to cover the entire atmosphere (Liu et al., 2018).

While these three dimensional models serve well to interpret satellite observations and study various atmospheric phenomena, their complexity also hinders the ability to test new ideas quickly and obtain a first order understanding of changes introduced by new theories. This gap in analytical tools is served by simpler models that reduce the order of the problem by accounting for fewer spatial dimensions. For example, Chamberlin and Hunten (1990) noted that while an atmosphere driven by solar inputs cannot be considered to be spherically symmetric, developing an understanding of a mean planetary atmosphere where day/night and latitudinal variations occur about the calculated mean values still produce useful insights. Such models are valuable in practice as well, having been used to test hypothetical solar spectra, variations in physical processes and new chemistry schemes (Smithtro and Sojka, 2005; Qian et al., 2006; Roble, 1995; Swaminathan et al., 2001).

In this paper we present the Atmospheric Chemistry and Energetics (ACE) 1D model, a first principles model capable of self-consistently solving for a mean thermosphere-ionosphere system. The model improves upon the work of Smithtro and Sojka (2005) and Roble (1987) by eliminating many prior assumptions, and incorporating an updated chemistry scheme reflecting our current understanding of aeronomic processes. Significant attention is given to detailing the chemistry and energetics of nitric oxide (NO), which plays an important role in governing the structure and temperature of the thermosphere despite being a minor constituent (Bailey et al., 2002; Mlynczak et al., 2003). The ACE1D model is the first to incorporate an updated and comprehensive NO chemistry scheme (Yonker, 2013; Yonker and Bailey, 2020), as well as incorporating NO chemiluminescence effects (Venkataramani et al., 2016). In particular, the chemistry scheme implemented by Yonker (2013) demonstrated that it is possible to reproduce equatorial NO densities at noon and above 100 km to within measurement uncertainties. In this paper we demonstrate its broader utility, extending it to calculate densities and temperatures of a globally averaged thermosphere and ionosphere. It is shown that the mean thermospheric and ionospheric profiles generated as a result of incorporating these advances retain agreement with established empirical models such as Mass-Spectrometer-Incoherent-Scatter (MSIS; Hedin, 1983) and the International Reference Ionosphere (IRI; Rawer et al., 1978), as well as measurements from the Global Ultraviolet Imager (GUVI) aboard the Thermosphere Ionosphere Mesosphere Energetics and Dynamics (TIMED) satellite (Yee, 2003) and the Student Nitric Oxide Explorer (SNOE) satellite mission.

The rest of the paper is laid out as follows: Section 2 of the paper provides a detailed description of the model. Section 3 discusses the model runs and results, while Section 4 compares results with observational data. Section 5 presents a discussion and conclusions from our work.

## 2. Model Description

The ACE 1D model produces a coupled, global mean thermosphere and ionosphere by self-consistently solving the continuity and energy equations and accounting for relevant chemical processes occurring in the upper atmosphere. The model assumes three major species: molecular nitrogen, molecular oxygen, and atomic oxygen, from which a number of neutral and charged minor species are produced due to the absorption of solar radiation by the atmosphere. The various functional components of the model and their interconnections are shown as a block diagram in Figure 1, and an overview of the model parameters is given in Table 1.

As shown in Figure 1, a specification of solar fluxes between 0.05-175 nm using the EUVAC model (section 2.8) forms the primary energy input to the model. Absorption of this energy by the major species ($N_2$, $O_2$, and O) leads to dissociation and ionization processes resulting in the formation of various neutral and charged minor species, as well as photoelectrons that drive secondary processes. The densities of major, minor and charged species are coupled via neutral-neutral and ion-neutral chemistry schemes. The energy released by these reactions, along with EUV absorption, Joule heating, and collisions with charged species serve as the energy source terms when solving for neutral temperatures. Radiative cooling by NO, $CO_2$ and $O(^3P)$ serve as the energy sinks for the neutral thermosphere. Lastly, a magnetospheric flux term is specified to serve as an energy input when calculating electron temperatures.

## 2.1 Pressure vertical co-ordinate

Dickinson and Ridley (1972) noted that physical distances are of little importance when expressing the continuity and heat equations in atmospheric models. The equations are significantly more tractable when using a pressure based coordinate system, yet retain a form that is applicable towards various planetary atmospheres. Following this work, the ACE1D model uses a pressure based coordinate system where the dependent variables are calculated on a vertical spatial grid defined as:

$$Z = -\ln\left(\frac{p}{P_0}\right) \qquad (1)$$

where $p$ is the pressure at the vertical grid point under consideration and $p_0$ is a reference pressure of $50\ \mu Pa$. A change of $\Delta Z = 1$ in this grid scheme corresponds to a change in altitude by one scale height, or a reduction in atmospheric pressure by a factor of $e$. In the current implementation, the model has a vertical resolution of $\Delta Z = 0.25$, with boundaries set at $Z = \pm 7$. The lower boundary of the model corresponds to a fixed altitude of 97 km, which is the approximate location of the mesopause averaged over a solar cycle (Smithtro and Sojka, 2005). The location of the model upper boundary ($Z = 7$) is dependent on solar activity and exospheric temperatures, and varies approximately between 450 and 750 km.

**Table 1.** ACE1D Model Parameters

| Major Species | $N_2$, $O_2$, O |
|---|---|
| Minor Species | NO, $N(^4S)$ , $N(^2D)$, $N(^2P)$, $N_2(A)$, $O(^1D)$, $CO_2$ |
| Ions | $O^+(^4S)$, $O^+(^2D)$, $O^+(^2P)$, $N^+$, $NO^+$, $N_2^+$, $O_2^+$ |
| Vertical Grid | $Z = -\ln\left(\frac{p}{p_0}\right), -7 \leq Z \leq +7,$ <br><br> $\Delta Z = 0.25, z_{lb} = 97\ km$ |
| Neutral Temperature | Lower boundary : Solar activity dependent, Upper <br><br> Boundary: $\frac{\partial T_n}{dZ} = 0$ |
| Election/Ion Temperatures | Lower boundary : $T_i = T_e = T_n$ <br><br> Upper Boundary : $\lambda \frac{\partial T_e}{\partial Z} = f$ |
| Solar Input | Solar Flux (0.05-175 nm), using EUVAC/F10.7 |
| Joule Heating Input | 70 GW Global Average |

The model assumes hydrostatic equilibrium, i.e., the pressure gradient force balances gravity:

$$\frac{dp}{dz} = -\rho g \qquad (2)$$

leading to a scale height definition of:

$$H = \frac{kT_n}{\bar{m}g} \qquad (3)$$

where $k$ is the Boltzmann constant, $T_n$ is the neutral temperature, $\bar{m}$ is the mean molecular mass, and $g$ is gravity.

## 2.2 Major species

The model calculates the relative density (or mass mixing ratio) defined for a species with molecular/atomic mass $m_i$ and number density $n_i$ as:

$$\Psi_i = \frac{m_i n_i}{\sum_j m_j n_j} \qquad (4)$$

where the summation in the denominator is over the three major species N₂, O₂, and O. The densities of O₂, and O are coupled in the thermosphere through dissociation and recombination reactions, and we define the vector $\mathbf{\Psi} = [\Psi_{O_2}, \Psi_O]$ which is solved for using the continuity equation:

$$\frac{\partial \mathbf{\Psi}}{\partial t} = -e^Z \, \tau^{-1} \frac{\partial}{\partial Z}\left[\frac{\bar{m}}{m_{N_2}}\left(\frac{T_{00}}{T}\right)^{0.25} \mathbf{\alpha}^{-1} \mathbf{L} \mathbf{\Psi}\right] + e^Z \frac{\partial}{\partial Z}\left(K_E \, e^{-Z}\left(\frac{\partial}{\partial Z} + \frac{1}{\bar{m}}\frac{\partial \bar{m}}{\partial Z}\right)\mathbf{\Psi}\right)$$
$$+ (\mathbf{S} - \mathbf{R}) \qquad (5)$$

where $\tau$ is the diffusion timescale characteristic of the thermosphere, $\bar{m}$ is the mean molecular mass, $T_{00}$ is a reference temperature equal to 273 K, and $\boldsymbol{S}$ and $\boldsymbol{R}$ are the coupled production and loss terms for O₂ and O. The above equation relates the time rate of change of the mass mixing ratios of O₂ and O to the gradient of the flux due to molecular and eddy diffusion (first and second terms) and the chemical production and loss of the species (third term). The matrix $\boldsymbol{\alpha}$ governs the diffusion of the major species, with elements defined as:

$$\alpha_{11} = -[\phi_{13} + (\phi_{12} - \phi_{13})\psi_2] \qquad (6)$$

$$\alpha_{22} = -[\phi_{23} + (\phi_{21} - \phi_{23})\psi_1]$$

$$\alpha_{12} = (\phi_{12} - \phi_{13})\psi_1$$

$$\alpha_{21} = (\phi_{21} - \phi_{23})2$$

with the subscripts 1, 2 and 3 referring to $O_2$, $O$ and $N_2$ respectively. $\phi_{ij}$ refers to the mutual diffusion coefficient, given as:

$$\phi_{ij} = \frac{D}{D_{ij}} \frac{m_3}{m_j} \qquad (7)$$

where $D$ is the characteristic diffusion coefficient:

$$D = D_0 \frac{P_{00}}{P} \left(\frac{P}{T_{00}}\right)^{1.75} \qquad (8)$$

$D_{ij}$ are the mutual diffusion coefficient for gases, with the values of $D_{12}$, $D_{13}$, $D_{23}$ equal to 0.26, 0.18, 0.26 respectively. $D_0$ is the characteristic diffusion coefficient at temperature $T_{00}$ (= 273 K) and pressure $P_{00}$ (= 105 Pa) and equals $0.2 \times 10^{-4}$ m$^2$ s$^{-1}$. $\boldsymbol{L}$ is a diagonal matrix operator with the elements:

$$L_{ij} = \delta_{ij} \left(\frac{\partial}{\partial Z} - \epsilon_{ii}\right) \qquad (9)$$

with

$$\epsilon_{ij} = 1 - \frac{m_i}{\overline{m}} - \frac{1}{\overline{m}} \frac{\partial \overline{m}}{\partial Z} \qquad (10)$$

The eddy diffusion coefficient $K_E$ is a prescribed parameter that is used to approximate turbulent mixing processes in the lower thermosphere and improve agreement between models and observational data. While $K_E$ does not have a definitive formulation, previous models have generally specified an altitude dependent value (Dickinson et al., 1984; Roble et al., 1987; Smithtro and Sojka, 2005), supplemented using harmonic functions to include seasonal effects (Qian et al., 2009). We use the expression given by Roble et al. (1987) which has units of $s^{-1}$:

$$K_E(Z) = 5 \times 10^{-6} \exp(-7 - Z) \qquad (11)$$

In order to solve the continuity equation, we constrain it by boundary conditions that reflect our understanding and behaviour of the thermosphere. While the models of Roble et al. (1987) and Smithtro and Sojka (2005) specify fixed number densities for the major species at the lower boundary, analysis of data from MSIS (Picone et al., 2002) suggests that the $O/O_2$ ratio at 97 km is not a constant. Rather, it increases with solar activity due to increased dissociation of $O_2$ into O. As a result we constrain Eq. 5 at the lower boundary by specifying $\Psi_{O_2}$ and $\Psi_O$ as solar activity dependent values obtained from MSIS. At the upper boundary we assume diffusive equilibrium and use the condition $\mathbf{L\Psi} = 0$. The ACE1D model employs centered spatial derivatives and forward time derivatives to discretize the continuity equation, and the resulting system of equation is solved using a tridiagonal matrix algorithm for the vector $\mathbf{\Psi}$. Finally, we obtain the mass mixing ratio for $N_2$ using the relation:

$$\Psi_{N_2} = 1 - \Psi_{O_2} - \Psi_O \qquad (12)$$

## 2.3 Minor species

The model calculates the densities of odd nitrogen species – NO, $N(^4S)$, $N(^2D)$, $N(^2P)$, the first excited state of molecular nitrogen, $N_2(A)$, and excited atomic oxygen, $O(^1D)$. The densities for most of these species are considered to be in photochemical equilibrium (PCE), given that their chemical lifetimes are short in comparison to the timescales for vertical transport. The densities in such cases are obtained as the ratio of the net production rate and loss frequency:

$$n_i = \frac{P_i}{L_i} \qquad (13)$$

Given their relatively longer lifetimes in the thermosphere, the model takes into account diffusion effects when calculating NO and $N(^4S)$ densities. The model solves for their relative densities using a form of the continuity equation similar to that used for the major species:

$$\frac{\partial \psi_i}{\partial t} = -e^Z \frac{\partial}{\partial Z} A_m \left[ \frac{\partial}{\partial Z} - E_i \right] \psi_i + e^Z \frac{\partial}{\partial Z} e^{-Z} K_E(Z) \left[ \frac{\partial}{\partial Z} + \frac{1}{\overline{m}} \frac{\partial \overline{m}}{\partial Z} \right] + (S_i - R_i) \qquad (14)$$

where

$$E_i = \left[ 1 - \frac{m_m}{\overline{m}} - \frac{1}{\overline{m}} \frac{\partial \overline{m}}{\partial Z} \right] - \alpha_i \frac{1}{T_n} \frac{\partial T_n}{\partial Z} + \overline{F} \qquad (15)$$

Once again, the first two terms of the continuity equation describe the gradient of the flux due to molecular and eddy diffusion processes, and the third term refers to chemical sources and sinks of the species being considered. $\alpha_i$ denotes the thermal diffusion coefficient, and $\overline{F}$ accounts for the frictional interaction between the minor species and the background atmosphere. Following the work of Colegrove et al. (1966), we set the thermal diffusion coefficient to be zero when solving for NO and N($^4$S). At the upper boundary, the model assumes diffusive equilibrium for both species, while at the lower boundary we specify a small downward flux for NO to account for transport into the lower atmosphere, and assume PCE for N($^4$S).

Production of $N_2$(A) in the thermosphere has been shown to play a significant role in nitric oxide chemistry (Yonker, 2013; Yonker and Bailey, 2020) by directly and indirectly influencing the production rates of NO and N($^2$D). Its densities are calculated by considering vibrational level specific production rates, obtained by scaling the photoelectron ionization rate of $N_2$, and considering losses to atomic oxygen and radiative relaxation in the Vegard-Kaplan bands. O($^1$D) is assumed to be produced via photodissociation of $O_2$ (Lee et al., 1977) and recombination of $O_2^+$ (Schunk and Nagy, 2009), and is lost to quenching by the neutral background atmosphere and via radiative relaxation at 630 nm. The model also includes the presence of $CO_2$ in the thermosphere as an inert constituent, as it is an important source of radiative cooling in the lower thermosphere. We specify a relative density value of $3.5\times10^{-4}$ at the lower boundary, and assume an e-folding height given by $H_{CO_2} = kT_n/m_{CO_2}g$.

## 2.4 Ionospheric species

ACE1D calculates the production of ionospheric species by accounting for direct photoionization and dissociative ionization of neutral species, ionization due to energetic photoelectrons, and charge exchange processes. Charge exchange processes also serve as the dominant loss mechanism for ions, except for NO$^+$ which generally forms the terminal ion in these reactions. NO$^+$ is lost only via dissociative recombination, which results in the production of N($^4$S) and N($^2$D) and is an important component of the NO chemistry (Swaminathan et al., 2001). The model assumes net charge neutrality and calculates the electron density to be equal to the sum of the ion densities.

Densities of ionospheric species are calculated primarily by assuming PCE. However, $O^+(^4S)$ and $N^+$ are assumed to undergo transport by diffusion, and their densities are calculated using the continuity equation (Schunk and Nagy, 2009):

$$\frac{\partial n_i}{\partial t} = (S_i - R_i n_i) - \frac{\partial \phi_i}{\partial Z} \qquad (16)$$

where $S_i$ and $R_i$ are the net production rate and loss frequency for a given species. The flux term $\phi_i$ is expressed as:

$$\phi_i = -\sin^2 I \ D_a \left( \frac{\partial n_i}{\partial Z} \frac{n_i}{H_p} + \frac{n_i}{T_p} \frac{\partial T_p}{\partial Z} \right) \qquad (17)$$

Where $D_a$ is the ambipolar diffusion coefficient, and $H_p$ and $T_p$ are the plasma scale height and temperature respectively. Ion densities are well represented by photochemical equilibrium at altitudes where eddy diffusion is important and is not included in the above equation. The magnetic dip angle $I$ denotes the orientation of the local magnetic field with the horizontal, and dictates the effectiveness of the field aligned diffusion in the vertical direction. A value of 0º corresponds to the direction of the field near the magnetic equator, where the field is parallel to the ground and there is no diffusion in the vertical direction. Similarly, 90º corresponds to the condition at the magnetic poles. The magnetic dip angle does not have a well defined value in the context of a global average thermospheric model (Smithtro, 2004), and we presently use a value of 75º to account for the deviation from a perfectly vertical magnetic field.

**2.5 Photoelectrons**

Photoelectrons are generated as a consequence of ionization of neutral species in the thermosphere by solar EUV radiation and soft X-rays. When the energy of the incident radiation exceeds the ionization potential of the absorbing species, the additional energy is converted into kinetic energy in the photoelectron, which may in turn continue to ionize, dissociate, excite and impart energy to other atoms and molecules.

As the production of photoelectrons are tied to a portion of the solar spectrum that varies significantly over various time scales, accurately accounting for their presence is an important aspect of modeling the thermosphere. A number of methods have been devised to estimate the effects of thermospheric photoelectrons – applying a scaling factor to the calculated total photoionization rate (Roble et al., 1987), parameterizing the photoelectron cascade calculations to approximate fluxes (Richards and Torr, 1983; Smithtro and Sojka, 2005), and obtaining a full numerical solution for the fluxes by solving electron transport equations (Bailey et al., 2002).

A more recent approach that has been implemented in global thermospheric-ionospheric models involves multiplying the wavelength dependent photoionization rate by appropriate scaling factors (Solomon and Qian, 2005) to account for the secondary ionizations, dissociative ionizations and dissociations caused by photoelectrons. This method avoids the need to calculate photoelectron fluxes themselves and reduces computational complexity, and has been shown to agree well with the full numerical solution obtained from the GLOW model of Solomon et al. (1988). ACE1D uses this parameterization scheme to account for photoelectron effects in the thermosphere. Collisions with photoelectrons also serve as the primary heat source for ambient ionospheric electrons, and thus have to be accounted for in electron temperature calculations. A parameterization for the volume heating rate of electrons due to photoelectrons between 120 – 325 km was initially given by Swartz and Nisbet (1972), where the heating rate was expressed as the product of a heating efficiency factor and the local photoionization rate. The heating efficiency factor in turn was expressed as a polynomial fit to a deposition parameter, a measure of the relative concentration of electrons to the neutral background atmosphere. An improved polynomial fit was subsequently given by Smithtro and Solomon (2008) that accurately reproduced the heating rate between 100-750 km. A further improvement of the parameterization was also given by Smithtro and Solomon (2008), where the photoelectron heating rates were calculated using separate fits for two broad band bins between 0-55 nm and 55-105 nm, which were defined on the basis of overall heating efficiency and variability under active solar conditions. This parameterization scheme has been shown to reproduce the electron heating rates for nominal and solar flare conditions within 15% of those calculated using the GLOW model (Smithtro and Solomon, 2008), provided that the solar flux below 5 nm is well specified.

The fluxes used by ACE1D meets this criteria, and thus implements the Smithtro and Solomon (2008) method to calculate the photoelectron heating rate of the ambient ionospheric electrons. Expressions for the heating efficiency factor, the deposition parameter and the polynomial fit coefficients used to calculate the heating rates are provided as supplemental information to this paper.

## 2.6 Neutral Gas Energy Equations

Neutral temperatures are solved for in the model using the one-dimensional time-dependent heat equation which accounts for molecular diffusion, eddy diffusion, and neutral heating and cooling processes:

$$\frac{\partial T_n}{\partial \mathrm{t}} = \frac{g}{P_0}\frac{e^Z}{C_p}\frac{\partial}{\partial Z}\left(\frac{K_T}{H}\frac{\partial T_n}{\partial Z} + K_E H^2 C_p \rho \left(\frac{g}{C_p} + \frac{1}{H}\frac{\partial T_n}{\partial Z}\right)\right) + \frac{Q-L}{C_p} \qquad (18)$$

Here, $C_p$ is the specific heat per unit mass, $H$ refers to the scale height, $K_T$ and $K_E$ are the thermal and eddy diffusion coefficients and $\rho$ is the average mass density. $Q$ and $L$ are the heat sources and sinks which are expanded upon in the following section. $P_0$ and $g$ are the model reference pressure of 50 µPa and the acceleration due to gravity respectively.

The expressions for specific heat capacity and thermal conductivity are obtained from Banks and Kockarts (1973):

$$C_P = k\left(\frac{7}{2}\left(\frac{v_{N_2}}{m_{N_2}} + \frac{v_{O_2}}{m_{O_2}}\right) + \frac{5}{2}\frac{v_O}{m_O}\right) \qquad (19)$$

$$K_T = \left(56\left(v_{O_2} + v_{N_2}\right) + 75.9\, v_O\right)T_n^{0.69}$$

Here $v_i$ and $m_i$ refer to the volume mixing ratio and molecular/atomic masses of the species, and $k$ is Boltzmann's constant. The eddy diffusion coefficient specification previously described in regards to the continuity equations remains the same.

The temperature at the lower boundary of the model is fixed to a global average value obtained from MSIS, and is dependent on solar activity. The average temperature at the lower boundary is 182 K, with a 1% variation over a solar cycle. At the upper boundary, thermal conduction dominates as the heat transport process and we specify the condition $\partial T_n / \partial Z = 0$.

## 2.7 Neutral temperature heating sources and sinks

Neutral gas heating in the thermosphere occurs via a number of processes, namely, 1) absorption of solar radiation by $O_2$ in the Schumann-Runge bands and continuum, 2) Joule heating, 3) exothermic neutral-neutral reactions 4) exothermic ion-neutral and ion recombination reactions, 5) quenching of excited species and 6) thermal collisions with electrons and ions.

A parameterized expression for heating due to absorption of solar radiation by $O_2$ in the Schumann Runge bands (175-205 nm) has been given by Strobel (1978), which is accurate to within 5% of measurements above 75 km. Similarly, an expression for the heating due to absorption in the Schumann-Runge continuum (125-175 nm) has been given by DeMajistre et al. (2001), which reproduces the heating rate within 2%, with uncertainty introduced primarily by the usage of the F10.7 index as a proxy for solar activity. However, the full calculation is not computationally expensive to implement for the low resolution solar flux specification used in the ACE model, and is obtained as:

$$E_C = E_{min} - (1 - \phi)E_{630} \qquad (20)$$

where $E_C$, $E_{min}$ and $E_{630}$ refer to the energy required to dissociate the $O_2$ molecule, the minimum energy required to excite one of the product oxygen atoms to the $^1$D level, and the energy difference between $O(^1D)$ and $O(^3P)$ respectively. $\phi$ is the wavelength dependent yield of $O(^1D)$ from the dissociation of $O_2$, obtained from Lee et al. (1977). The heating rate is then calculated by summing the differences between $E_c$ and the energy of the absorbed photon across all wavelength bins.

Previous modeling efforts by Roble et al. (1987) and Smithtro and Sojka (2005) have noted the need to include joule heating in order to bring the calculated global average neutral temperatures into agreement with results from MSIS. In both these models, a global energy input of 70 GW proposed by Foster et al. (1983) for geomagnetically quiet conditions was used. More recently, Knipp et al. (2004) presented a value of 95 GW ± 93 GW for the daily average joule heating power, averaged for solar cycles 21 - 23. Due to the large variability present in this value, we continue to use a value of 70 GW here and consider the sensitivity of the ACE1D model to changes in the joule energy input in future studies. The joule heating rate is calculated as the product of the Pederson conductivity and the superimposed electric field:

$$Q_{Joule} = \sigma_P \, E^2 \qquad (21)$$

We assume the electric field to vary between 4.07-6.74 mV m$^{-1}$ to keep the net energy input constant.

The ACE1D model considers infrared emissions at by NO, $CO_2$ and O($^3$P) as sources of radiative cooling in the thermosphere. Both NO and $CO_2$ convert kinetic energy into vibrational energy, which subsequently gets radiated at 5.3 and 15 μm respectively. In comparison, the 63 μm emission by O($^3$P) occurs solely due to atomic magnetic dipole transitions (Kockarts and Peetermans, 1970; Bates, 1951).

The radiative cooling rate by nitric oxide is given as

$$L_{NO} = h\nu \, A_{10} \, [NO_{v=0}] \, \omega \, e^{-\frac{h\nu}{kT_n}} \qquad (22)$$

where

$$\omega = \frac{k_{10}[O]}{k_{10}[O] + A_{10}} \qquad (23)$$

Here, $h\nu$ is the energy of the emitted photon, $A_{10}$ is the Einstein coefficient, $k_{10}$ is the rate coefficient for collisional quenching of vibrationally excited NO by O, and the quantities in square brackets denote number densities of the species. Several values of $k_{10}$ have been presented and used over the years in aeronomic calculations, with Kockarts (1980) and Roble et al. (1987) using a value of $6.5 \times 10^{-11}$ cm$^3$ s$^{-1}$, Sharma et al. (1998) using a value of $2.8 \times 10^{-11}$ cm$^3$ s$^{-1}$, and Lu et al. (2010) using a value of $4.2 \times 10^{-11}$ cm$^3$ s$^{-1}$. A contextual review of the various rate coefficients and the models in which they were used is given in Venkataramani (2018). In the present model, we use a value of $2.1 \times 10^{-11}$ cm$^3$ s$^{-1}$ as calculated by Caridade et al. (2008). A goal of this paper is to determine if this rate coefficient can produce a reasonable modeled atmosphere A full calculation of the chemiluminescent emissions at 5.3 and 2.7 μm generated as a consequence of NO production of in the thermosphere (Venkataramani et al., 2016), is included in the model as well. However, we note that these emissions only reduce the effective exothermicity of the underlying reactions rather than serving as explicit thermospheric cooling mechanisms.

An expression for the cooling rate above 95 km due to the emission from CO$_2$ assuming non-local thermodynamic equilibrium (non-LTE) was given by Dickinson (1984). This expression is adjusted to account for the experimental results of Castle et al. (2012):

$$L_{CO_2} = h\nu \, n(M) \, n(CO_2) \, \lambda_{VT1} \times 2 \, e^{-\frac{960}{T_n}} \qquad (24)$$

where

$$\lambda_{VT1} = \lambda_{VT}^M + r_o \, \lambda_{VT}^o \qquad (25)$$

$$\lambda_{VT}^M = 2.5 \times 10^{-15} \, cm^3 \, s^{-1} \, ; \, T < 200 \, K$$
$$= 2.5 \times 10^{-15} \left( 1 + 0.03(T - 200) \right) cm^3 \, s^{-1}; \, T \geq 200 \, K$$

$$\lambda_{VT}^o = (3.8 - 9.51 \times 10^{-3} \, T_n + 9.32 \times 10^{-6} \, T_n^2) \times 10^{-12} \, cm^3 \, s^{-1} \, ; \, 142 \leq T_n \leq 490 \, K$$
$$= 1.38 \times 10^{-12} \, cm^3 \, s^{-1}; \, T_n > 490 \, K$$

An expression for the emission due to O($^3$P) Is given by Kockarts and Peetermans [1970], which also included an altitude dependent reduction factor that accounts for self-absorption and varies between 0.0 at 100 km and 0.8 at 200 km and above. Roble et al. (1987) scaled this cooling rate by a factor of 0.5 following the measurements of Offermann and Grossmann (1978) which suggested that the O($^3$P) population in the upper thermosphere was not in thermodynamic equilibrium (non-LTE conditions). Subsequent modeling work by Sharma et al. (1994) and measurements by Grossmann and Vollmann (1997) have shown this assumption to be incorrect, and we do not apply a scaling factor in our model. We use the following expression for the radiative cooling due to O($^3$P) in the model:

$$L_{O(^3P)} = X' \, \frac{1.69 \times 10^{-18} \, n(O) \, e^{-\frac{228}{T}}}{1 + 0.6 \, e^{-\frac{228}{T}} + 0.2 \, e^{-\frac{326}{T}}} \, ergs \, cm^3 \, s^{-1} \qquad (26)$$

Here, $X'$ is the altitude dependent reduction factor, and $n(O)$ is the density of atomic oxygen.

**2.8 Electron and Ion Energy Equations**

Compared to the background neutral atmosphere, the electron temperature responds quickly to changing conditions and can be solved for by assuming steady state conditions. The electron energy equation takes the form:

$$\sin^2 I \frac{\partial}{\partial Z} \left[ K_e \frac{\partial T_e}{\partial z} \right] + Q_e - L_e = 0 \qquad (27)$$

Where $I$ is the magnetic dip angle. An expression for the electron conductivity $K_e$ has been given by Rees and Roble (1975) which accounts for the deviation from that of a fully ionized plasma due to electron-neutral collisions:

$$K_e = 7.5\text{x}10^5 \, T_e^{5/2} \left( 1 + 3.22 \text{ x } 10^4 \frac{T_e^2}{n_e} \sum_{n=1}^{3} n_n \, \overline{Q_{D_n}} \right)^{-1} \qquad (28)$$

where $n_n$ refers to neutral densities and $\overline{Q_{D_n}}$ are the velocity averaged momentum transfer cross sections. The summation in the expression is applied over the densities of the major species.

Ambient ionospheric electrons are primarily heated by collisions with photoelectrons and are cooled by inelastic and elastic collisions with neutrals and ions. Expressions for the energy loss rates from electrons have been given by Schunk and Nagy (2009) and have been provided as supplemental information. Quenching of N($^2$D) is also known to be a significant source of electron heating in the F2 region of the ionosphere (Richards, 1986), and is included in the model calculations. It is necessary to specify a heat flux at the upper boundary corresponding to the energy input from the magnetosphere in order to obtain a reasonable electron temperature profile. We adopt a fixed value of $3 \times 10^9$ eV cm$^{-2}$ s$^{-1}$ following Roble et al. (1987) as the upper boundary condition in solving for the electron temperature.

The ion temperatures are solved for by assuming local thermodynamic equilibrium. The coulomb collisions that serve as a cooling mechanism for ionospheric electrons serves as the heating mechanism of the ions, while collisions with neutrals serves as the heat sink. Expressions for these energy loss rates have been reproduced from Rees and Roble (1975) in the supplemental document. The electron and ion temperatures are set to be equal to the neutral temperature at the lower boundary of the model.

## 2.9 Chemistry Scheme

The sources and sinks referred to in the density and energy equations detailed in the previous sections are governed by a chemistry scheme that is based on the work of Roble et al. (1987), but updated to reflect subsequent updates to reaction rates, product yields, exothermicities, and processes that have not been taken into account. The model utilizes the work of Yonker (2013) as the basis for updating these values, as well as the parameterization scheme used to calculate $N_2(A)$ densities which serves as a source for thermospheric NO.

## 2.10 Model Inputs

The ACE1D model uses a combination of solar fluxes, joule heating and magnetospheric flux specification as energy inputs to the thermosphere. Of these, solar fluxes are the dominant input which are varied a function of solar activity. Fluxes for a given level of solar activity is calculated by scaling the photon fluxes in individual wavelength bins, following the method used in the EUVAC model (Richards et al., 1994):

$$F(\lambda) = F_{ref}(\lambda)[1 + A(\lambda)(P - 80)] \qquad (29)$$

$$P = (F10.7_d + F10.7_a)/2$$

where $\lambda$ is the wavelength bin under consideration, $F_{ref}$ refers to the refence solar minimum spectra and $A$ is a wavelength dependent scaling factor. $P$ is a proxy for solar activity, calculated as the average of the daily ($F10.7_d$) and 81 day averages ($F10.7_a$) of the 10.7 cm solar flux index. The model is designed to reproduce the reference solar spectra for $P = 80$, and the fluxes are scaled to 80% of their reference value at all wavelengths for $P < 80$.

Figure 2 shows the solar fluxes used to drive the model at solar minimum and maximum ($P = 70, 250$), along with the reference solar spectrum to which the scaling factors are applied. To accurately reproduce the calculations obtained from high resolution solar spectra, the model uses variable bin sizes to specify the flux between $0.05 - 105$ nm, and 5 nm wide bins between 105 and 175 nm. A separate bin is used to specify the flux for the Lyman-α emission at 121.57 nm. The variability of the solar spectrum is described by the ratio $S_{max}/S_{min}$, which indicates significant variation in the EUV and XUV fluxes as a result of the variability present in the solar corona where they are produced. In comparison, fluxes at longer wavelengths of the solar spectrum are emitted by regions closer to the solar surface and exhibit less variability.

## 2.11 Globally Averaged Energy Inputs

In order to compute a globally averaged thermosphere where only one hemisphere is sunlit at any given time, the solar irradiance input to the model is scaled by a factor of 1/2 across all wavelengths. Further, it is necessary to specify an average value for the solar zenith angle to account for the local time variation on the sunlit side. Using the expression for solar zenith angle (Jacobson, 2005):

$$cos\,\theta_s = sin\,\phi\,sin\,\delta + cos\,\phi\,cos\,\delta\,cos\,H_a \qquad (30)$$

Here $\theta_s$, $\phi$, $\delta$ and $H_a$ refer to the solar zenith angle, latitude, solar declination angle and local hour angle of the Sun, respectively. Then, the area weighted average of the cosine of the solar zenith angle for the sunlit half of the Earth is evaluated as:

$$< cos\,\theta_s > = \frac{1}{2\pi r^2} \int_{-\frac{\pi}{2}}^{\frac{\pi}{2}} \int_{-\frac{\pi}{2}}^{\frac{\pi}{2}} r^2 \, cos^2\,\phi\,cos\,H_a \; d\phi \, dH_a \qquad (31)$$

where we take advantage of the spherical symmetry and simplify the expression to integrate about the subsolar point ($\delta = 0$). Here $r$ refers to the Earth's radius, and we integrate between $-\pi/2$ and $\pi/2$ for the local solar angle ($H_a = 0$ refers to local noon). Evaluating this expression, we obtain the globally averaged value for the solar zenith angle as $\theta_s = 60°$.

## 2.12 Initial Conditions

The model is initialized to a reference atmosphere comprised of the major species: O, $O_2$, $N_2$; minor species: NO, N($^4$S), N($^2$D), and charged species: $O^+$($^4$S) and $e^-$. The remaining neutral and charged species are initialized to zero values. The neutral, ion, and electron temperatures are also initialized to reference values. The model is then run at a 60 second timestep for 30 days, allowing it to converge upon a steady state solution.

# 3 Model Results

In order to test the ability of the model to generate a reasonable representation of the thermosphere/ionosphere system, multiple model runs were conducted with varying solar fluxes generated by the EUVAC model by varying $P$ from a value of 70 (solar minimum) to 250 (solar maximum). Previous modeling efforts have noted the need to include additional energy inputs to obtain a reasonable thermospheric neutral temperature profile. To facilitate comparisons with these studies, we use a joule heating input of 70 GW in the neutral energy equation, and a magnetospheric heat flux of $3 \times 10^9$ eV cm$^{-2}$ s$^{-1}$ in the electron energy equation.

Figures 3 and 4 present an overall view of the model outputs for the two extreme cases of model runs, and Figures 5 and 6 provide additional information on the behaviour of exospheric temperatures and peak NO densities as calculated by the model. The variability of varying heating and cooling processes over the course of a solar cycle is shown in Figures 7 and 8. Comparisons of model results with GUVI and SNOE data are shown in Figures 6 and 9. In order to compare ACE1D results with MSIS, we generate binned MSIS data over a global grid for an entire solar cycle, and then compute the area weighted averages for temperatures and densities.

## 3.1 Major Species

The model results for the densities of major species are shown in Figures 3(a) and 4(a). The calculated densities are in reasonable agreement with MSIS values, with the largest difference between the two models seen in the $O_2$ densities at solar maximum near 170 km, where the

ACE1D model predicts a value of $3.25 \times 10^8$ cm$^{-3}$ compared to the MSIS value of $2.22 \times 10^8$ cm$^{-3}$ (46% difference). The calculated O densities have a maximum discrepancy of 30% during solar minimum, with ACE1D calculating a value of $1.09 \times 10^{11}$ cm$^{-3}$ at 115 km compared to the MSIS value of $1.59 \times 10^{11}$ cm$^{-3}$. At solar maximum the discrepancy is about 25%, where ACE1D calculates a value of $1.1 \times 10^{11}$ compared to the MSIS value of $1.48 \times 10^{11}$ cm$^{-3}$.

## 3.2 Minor Species

Figures 3(b) and 4(b) show the calculated densities of neutral minor species. As stated in section 2.3, the model assumes photochemical equilibrium for all species except NO and N($^4$S), for which vertical transport and eddy diffusion effects are included.

The ACE1D model calculates the v = 0 − 7 level densities for N$_2$(A), and the sum of these densities are shown in Figure 3(b) and 4(b). For equatorial noon, Yonker (2013) calculated a peak density of approximately 2.95–3.8$\times 10^3$ cm$^{-3}$ near 160 km for moderately high levels of solar activity ($P = 155, 183$). In comparison, the ACE1D model calculates a peak density of $5.6 \times 10^2$ –1.4$\times 10^3$ cm$^{-3}$ at solar minimum and maximum respectively, with a peak altitude ranging between 150 − 170 km. These differences are expected given the strong diurnal variation in the production rate of N$_2$(A) and the global average profile that the model generates.

The electronically excited state of atomic oxygen, O($^1$D), is the source of the 630 nm airglow emission from the thermosphere. It is primarily produced in the thermosphere via O$_2^+$ recombination, O$_2$ dissociation, and photoelectron excitation of ground state atomic oxygen. As photoelectron fluxes are not explicitly calculated in the model, excitation of atomic oxygen is not presently included as a source of O($^1$D) and results in significantly smaller calculated densities when compared to sounding rocket measurements (Solomon and Abreu, 1989). While the reaction of $N(^2D)$ with $O_2$ as a potential source of $O(^1D)$ has been discussed in the past (Solomon et al., 1988; Link and Swaminathan, 1992), the model follows the most recent experimental result of Miller and Hunter (2004) which indicates a negligible yield for the $O(^1D)$ channel. The model predicts a peak density of approximately $3.3$-$3.6 \times 10^3$ $cm^{-3}$ near $125 - 140$ km for solar maximum and minimum respectively.

Densities of the excited atomic nitrogen species $N(^2D)$ and $N(^2P)$ are inferred in the thermosphere by observing emissions at 520 and 346.6 nm, corresponding to their radiative relaxation to the $N(^4S)$ ground state. However, there are no definitive global measurements of these emissions. The model calculates a peak $N(^2D)$ density of $2 \times 10^5$ $cm^{-3}$ for solar minimum and $5.3 \times 10^5$ $cm^{-3}$ for solar maximum, while the corresponding values for $N(^2P)$ are $7.3 \times 10^2$ $cm^{-3}$ and $1.5 \times 10^3$ $cm^{-3}$. While the peak altitude for $N(^2D)$ densities calculated by the model agree with previous calculations by Roble (1987) and Smithro et. al (2005), the magnitude of the ACE1D modeled densities are smaller by approximately a factor of 2 when compared to the values of Roble (1987). This is attributed to the larger, temperature dependent reaction rate prescribed by Duff et al. (2003) used in our model, which results in increased loss to $O_2$ and a corresponding increase in NO densities.

The NO chemistry implemented in the ACE1D model follows the comprehensive review of processes given by Yonker (2013). Figure 3(b) and 4(b) show ACE NO densities calculated assuming PCE, as well as when including diffusion effects. It can be seen from these figures that diffusion has a negligible effect on the global mean peak NO densities near 110 km, particularly for high solar activity. The peak density for the two cases are 3.7×10$^7$ cm$^{-3}$ and 9.6×10$^7$ cm$^{-3}$ respectively, which is a factor of 2 – 3 larger than those predicted by Roble et al. (1987). Figure 5 shows the NO profile between 100 and 200 km calculated for $P = 70, 150$ and 200, which indicates a slight decrease in the location of the density peak from 109 to 106 km with increasing solar activity, caused by the increase in the production from N($^2$D). The location of the peak was however found to be independent of the downward flux specified at the lower boundary to constrain Eq. 14.

Given its longer lifetime, the effects of vertical diffusion is significantly stronger on N($^4$S) densities as compared to NO, with a noticeable shift in peak altitude when comparing PCE based density calculations with those where diffusion effects are included. In both cases, the peak altitude is lower by approximately 100 km when diffusion effects are included, and the peak densities are reduced by a factor of 5. The model calculates a peak density of 1.5x10$^7$ and 5x10$^7$ cm$^{-3}$ at solar minimum and solar maximum, which are in good agreement with the values presented by Roble (1987).

**3.3 Neutral, Electron and Ion Temperatures**

The neutral, electron and ion temperatures calculated by the model are shown in Figure 3(c) and 4(c), along with neutral temperatures from MSIS. We find that the model underestimates thermospheric neutral temperatures when provided solar fluxes as the only energy input, producing exospheric neutral temperatures of 546 and 1271 K at solar minimum and maximum respectively. While the model outputs are in reasonable agreement with MSIS at higher levels of solar activity, it was found necessary to include a joule energy input in the model in order to bring the temperatures into reasonable agreement with MSIS over a wide range of solar activity.

We thus specify a global joule energy input of 70 GW, resulting in model calculated exospheric neutral temperatures of 643 K and 1364 K compared to MSIS values of 740 and 1263 K, corresponding to a 8-13% difference. Figure 6 plots the exospheric temperatures from ACE1D and MSIS as a function of solar activity, which show agreement within 15% over the course of a solar cycle. The figure also plots daytime exospheric temperatures obtained from GUVI (Meier et al., 2015), which made limb observations between 2002-2007 on the dayside (local time 12-18 hours) between ±55°, for moderate levels geomagnetic activity ($A_p$ = 5-6). Similar to the MSIS data, we bin the GUVI temperatures by latitude, longitude, and solar activity, and compute each data point as an area weighted average. The temperatures are plotted as a function of the solar activity proxy value $P$ (section 2.8). It should be noted that GUVI observations at high values of solar activity ($P > 200$) occurred primarily at large zenith angles during dawn, resulting in significantly lower exospheric temperatures.

While the GUVI observations cannot be compared directly with the ACE1D results, both MSIS and GUVI indicate a non-linear relationship with solar activity that the model is unable to reproduce, likely caused by inadequate specification in the solar input. Figure 6 also shows the integrated energy from XUV/EUV solar fluxes between 0.05 - 110 nm, which while well correlated with the calculated exospheric temperatures, is also modeled to be a linear function of solar activity. The model calculations thus indicate that the scaling applied to calculate variations in the solar fluxes do not adequately capture the actual variability over the course of a solar cycle.

**3.4 Ion Densities**

Figure 3(d) and 4(d) shows model calculations for globally averaged ion and electron densities, along with globally averaged IRI values obtained for similar solar activity levels. The model calculates a peak electron density of $1.5 \times 10^5$ cm$^{-3}$ at 230 km for solar minimum and a density of $8.7 \times 10^5$ cm$^{-3}$ at 330 km for solar maximum. This is smaller than the IRI values of $3.1 \times 10^5$ cm$^{-3}$ (factor of 2) and $1.2 \times 10^6$ cm$^{-3}$ (factor of 1.4) at solar min and max respectively. The E region of the ionosphere below 200 km is comprised primarily of $NO^+$ and $O_2^+$ ions, while $O^+$ is the dominant ion in the F region above 200 km. $N_2^+$, $N^+$, $O^+(^2D)$ and $O^+(^2P)$ are seen to be minor ionospheric constituents at all altitudes, which are produced primarily by the ionization and dissociative ionization of $N_2$ and $O_2$.

**3.5 Heating and Cooling rates**

Model calculations of the various neutral gas heating sources are shown for solar minimum and maximum in Figure 7, along with the fractional contribution of each process as a function of altitude. The peak heating rates calculated for the two cases are $1.3 \times 10^5$ ergs g$^{-1}$ s$^{-1}$ and $5.2 \times 10^5$ ergs g$^{-1}$ s$^{-1}$ respectively, compared to the values of $1.5 \times 10^5$ ergs g$^{-1}$ s$^{-1}$ and $4.6 \times 10^5$ ergs g$^{-1}$ s$^{-1}$ obtained by Roble et al. (1987).

Thermospheric heating can broadly be considered to be dominated by processes involving neutral species below 200 km and charged species above this altitude. The importance of individual processes to the net heating rate is also strongly dependent on altitude and solar activity. It is seen that absorption at EUV wavelengths by $O_2$ in the Schumann Runge continuum and bands, and quenching of excited species are the dominant heating sources below 150 km.

Absorption in the Schumann Runge continuum and bands has a peak contribution to the net heating rate that varies between 50% at solar minimum to 36% at solar maximum, while that due to quenching of excited species remains constant at approximately 55%. Exothermic neutral-neutral reactions are important near the base of the thermosphere where atomic oxygen recombination occurs, and between 150-200 km where heating due to odd nitrogen chemistry is important. Joule heating, which is specified to have a fixed power input of 70 GW in the model, decreases in importance as a heating mechanism with increasing solar activity. At solar minimum it has a peak contribution of 27% in the lower thermosphere, compared to 15% at solar maximum. Heating due to collisions with energetic photoelectrons has a peak contribution of approximately 10% for both levels of solar activity. The list of exothermic neutral-neutral reactions implemented in the model have been tabulated and provided as supplemental information along with their respective rate coefficients and exothermicities. The recombination reaction of atomic oxygen is the dominant source of heating at the lower boundary of the model, while the quenching of excited species is important above 110 km. The exothermic reactions of $N(^2D)$ and $N(^4S)$ with $O_2$ are important near 150 km, but their efficiency is significantly reduced due to the chemiluminescence of NO discussed previously. We assume that the heating rate of neutrals due to direct impact with photoelectrons is 5% of the total EUV energy absorbed (Stolarski, 1976; Roble et al., 1987). Energy exchange via collisions that serve as the heat sink for electrons and ions serve as the primary heat input into neutrals in the upper thermosphere.

The components of thermospheric cooling for solar minimum and maximum are shown in Figure 8. It is important to note that molecular conduction and eddy diffusion are not strictly "cooling" processes, as they generally serve to transport heat from higher to lower altitudes. This was also observed by Bates (1956) who noted that thermal conductivity is independent of gas density, but the rates of heat absorption and loss reduce with gas density. This means that conduction is an important contributor in the thermal balance of the upper atmosphere and is reflected in the dominant role played by heat conduction above 150 km in transporting energy downward during both solar minimum and maximum. This transition from heat sink to a heat source is reflected in the sharp inflection points seen in the thermal conduction cooling rate near 120 km during solar minimum, and 150 km at solar maximum.

Emissions at 15 μm from $CO_2$ are the dominant source of radiative cooling below 120 km, while the 63 μm emission from $O(^3P)$ is important above 200 km. At solar minimum, the contribution of the 5.3 μm emission from NO to the net cooling rate is smaller than either of the above emissions. However, while the energy loss rates due to $CO_2$ and $O(^3P)$ are seen to increase by approximately 33% and 12% respectively between solar minimum and maximum, the peak cooling rate due to NO increases by nearly an order of magnitude, with the peak emission increasing from $3.2×10^3$ ergs g$^{-1}$ s$^{-1}$ to $3×10^4$ ergs g$^{-1}$ s$^{-1}$. This increase results from a combination of increases in the NO densities and the temperature of the thermosphere, and serves to highlight the "thermostat" effect of the 5.3 μm cooling [Mlynczak, 2003].

Figure 8 also shows the contribution of the chemiluminescent emission from nitric oxide to energy loss from the thermosphere. As previously noted in section 2.5, this emission is not a cooling source in the traditional sense, as it does not convert kinetic energy from the background atmosphere into infrared radiation. Instead, it reduces the effective exothermicities of the reactions that produce nitric oxide in the thermosphere. With this consideration, it is seen that emissions from nitric oxide due to chemiluminescence has a peak value of $5.3 × 10^3$ ergs g$^{-1}$ s$^{-1}$ at solar minimum and are comparable to the collisional emission. The fractional contribution of chemiluminescent processes to NO emissions at solar maximum is approximately 45%, with a peak value of $1.3 \times 10^4$ ergs g$^{-1}$ s$^{-1}$.

## 4   NO and Model temperature sensitivity

A primary motivation for developing the ACE1D model was to enable testing of changing rate coefficients etc. in models. We conduct a few experiments here in order to test the effect of nitric oxide cooling on the thermospheric energy budget and temperatures.

### 4.1 The effect of NO chemiluminescence

Vibrational excitation of NO in the thermosphere occurs via either collisional excitation with atomic oxygen, or as a consequence of the exothermic reactions that produce it. Understanding the NO chemiluminescent emission is important as it contributes to thermospheric emissions at 5.3 μm (Mlynczak, 2003) but differs in the role it plays in thermospheric energetics. Since collisional excitation of NO to higher vibrational levels (v>1) is unlikely, chemiluminescence also serves as a unique marker for thermospheric NO production rates which can be obtained by studying the overtone emission (Δv=2) at 2.7 μm (Wise et al., 2001). Calculating level populations and the subsequent emissions for NO(1≤v≤10) via both these mechanisms is detailed in Venkataramani (2016), and has been implemented in the present model. The effect of accounting for this phenomenon is shown in Figure 9, which compares neutral temperatures obtained for various model runs at $P = 70$ and 250 with MSIS results.

Including chemiluminescent NO emissions reduces the contribution of neutral heating at altitudes where NO is produced, which is primarily below 200 km. This reduction is partially offset by downward molecular heat conduction (Figure 8), resulting in a relatively unchanged temperature structure below 150 km. Including this mechanism thus primarily affects the isothermal region of the thermosphere, reducing exospheric temperatures from 665 to 643 K at solar minimum and from 1406 K to 1364 K at solar maximum. Figure 9(b) shows the effect over the course of a solar cycle, denotes a nearly uniform 3% reduction in exospheric temperatures.

## 4.2 NO cooling rates

The rate coefficient for the vibrational excitation of NO by collisions with atomic oxygen is generally determined by studying the reverse reaction, i.e. the collisional quenching of NO(v = 1) by atomic oxygen. Several values have been presented and used over the years for aeronomic calculations, with Kockarts (1980) and Roble et al. (1987) using a value of $6.5\times10^{-11}$ cm$^3$ s$^{-1}$, Sharma et al. (1998) using a value of $2.8\times10^{-11}$ cm$^3$ s$^{-1}$, and Lu et al. (2010) using a value of $4.2\times10^{-11}$ cm$^3$ s$^{-1}$.

Current thermospheric models (Lu et al., 2010) generally use the experimental results given by Hwang et al. (2003), which is potentially an overestimate (Sharma and Roble, 2002; Venkataramani, 2018) . While the ACE1D model currently uses the value of $2.1\times10^{-11}$ cm$^3$ s$^{-1}$ (Caridade et al., 2008), the results of model runs by doubling the NO cooling rate to the Hwang et al. (2003) value is presented in Figure 9. This leads to an exospheric temperature of 626 and 1280 K at $P = 70$ and 250. While this suggests a good agreement with the MSIS predicted value for solar maximum, Figure 9(b) shows that the agreement is valid only at higher levels of solar activity, while resulting in a 12-15% difference from MSIS value for $P = 70 - 150$.

# 5    Comparisons with SNOE data

The SNOE satellite mission (Barth et al., 2003) made measurements of NO densities in the thermosphere between 95 – 170 km from 1998 to 2003. Placed in a sun synchronous orbit, SNOE made near global measurements each day at an average local time of 11 am. Figure 10 shows global average SNOE measurements of NO density at 106 km plotted as a function of solar activity, along with the standard deviation of the measurements. The increase in the variability of the measurements with solar activity reflects the variability in the XUV and EUV solar fluxes that drive NO production in the thermosphere. The ACE1D calculations agree with the general trend of the measurements from SNOE, but they are approximately 25% smaller than the mean SNOE values at moderate to high solar activity levels. We expect this as the model calculates a global average value for the NO peak, while the local time of SNOE measurements varied between 10-11 am.

# 6    Discussion and conclusions

The ACE1D model is a first principles based thermospheric-ionospheric model that is capable of generating a self-consistent global mean vertical profile of the Earth's upper atmosphere. This is done by solving the continuity equation for the major species, including vertical transport effects, but ignoring horizontal winds. Neutral, ion, and electron temperatures are solved for using the appropriate heat equations. In both continuity and energy equations, altitude dependent eddy diffusion effects are taken into account by using parameterization schemes established in literature. The model employs an updated chemistry scheme that allows for the calculation of a number of minor neutral and ion species, assuming photochemical equilibrium where appropriate and including effects of vertical diffusion for species with longer lifetimes. Energy inputs to the model are specified in terms of a 37 bin solar spectra, parameterized using daily and 81 day average F10.7 index. Additionally, we specify a fixed joule heating input of 70 GW in the neutral energy equation and a magnetospheric heat flux as a heating source for ionospheric electrons. The model is the first to self-consistently incorporate the comprehensive NO chemistry scheme presented by Yonker and Bailey (2020), as well as include the effects of NO chemiluminescence. The ability of the model to reproduce a number of thermospheric features shows the broader applicability of this chemistry scheme in modeling the thermosphere and ionosphere.

Calculations from the model for neutral density and temperatures are in good agreement with MSIS values over a range of solar activity levels. The model results also compare reasonably with observations of thermospheric neutral temperature and NO density measurements from satellite missions. Discrepancies between calculated and empirically obtained exospheric temperatures are on average within 10%, and can be improved by better specification of solar fluxes and a re-examination of radiative cooling mechanisms in the thermosphere. We believe that improving the solar flux specification will also help reproduce the non-linear response of exospheric temperatures to increasing solar activity predicted by MSIS, which in contrast to the relatively linear response predicted by ACE1D and similar models.

It is notable that the ACE1D model calculated exospheric temperatures are on average within 15% of MSIS calculated values, particularly given the fact that it uses a rate coefficient for the 5.3 $\mu$m NO emission that is half of the value specified in prevalent thermospheric models, and that it includes effects of NO chemiluminescence. This suggests that a smaller rate coefficient for the collisional quenching of NO is capable of accounting for energetics in the thermosphere, provided that it is supported by a comprehensive NO chemistry scheme. At the very least, this suggests the need for further experimental studies of the NO(v=1) + O quenching process, which is presently limited to a few measurements (Dodd et al, 1999; Hwang et al., 2003).

While the model is limited in generating an average representation of the global thermosphere-ionosphere system and cannot reproduce specific observations, it serves as a valuable analytical tool to quickly evaluate new theories of upper atmospheric processes. Consequently, the ACE1D model can serve to test variations in modeling parameters and processes before they are implemented into more complex models of the thermosphere.

**Figures**

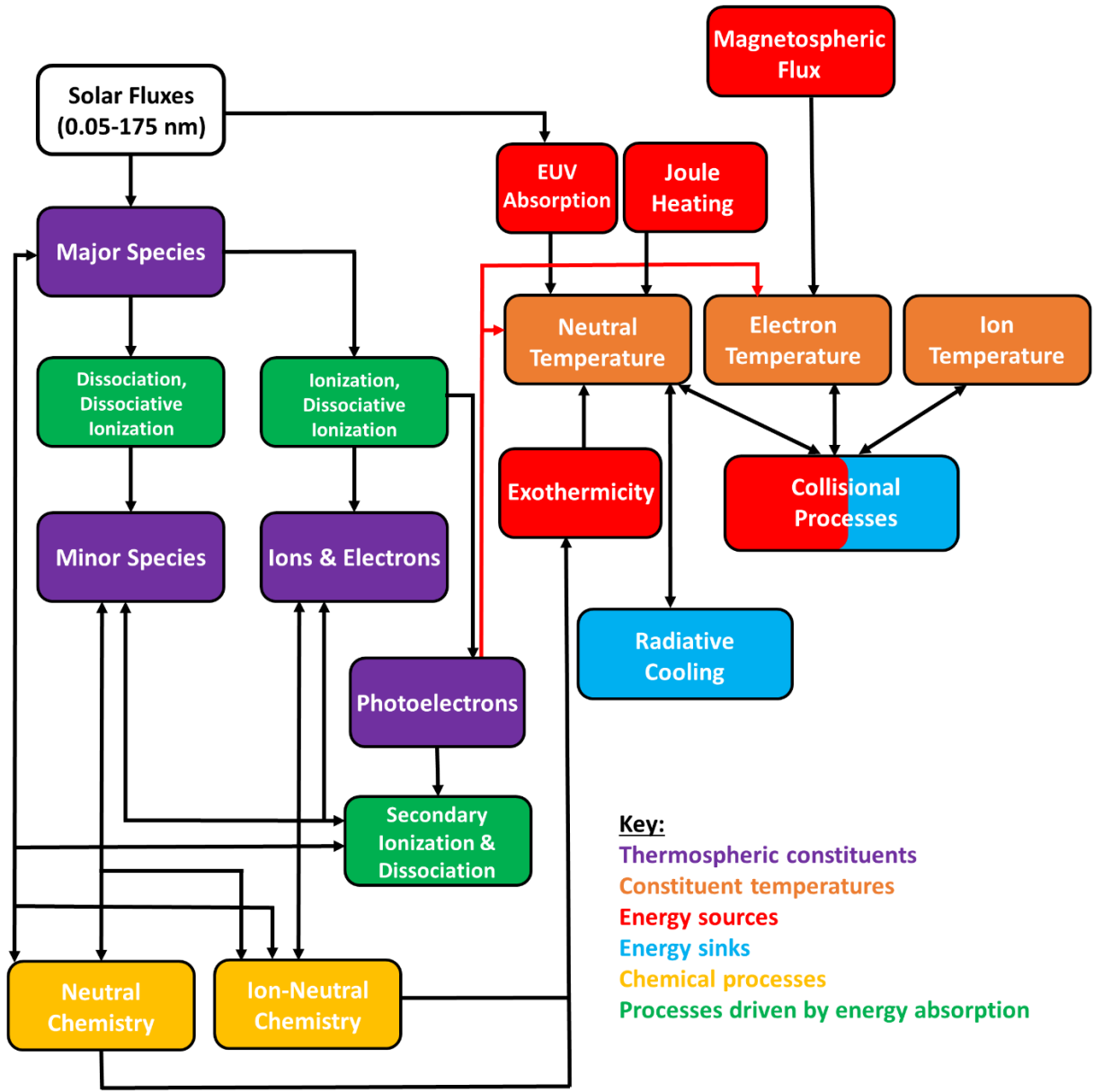

Figure 1: ACE1D model block diagram

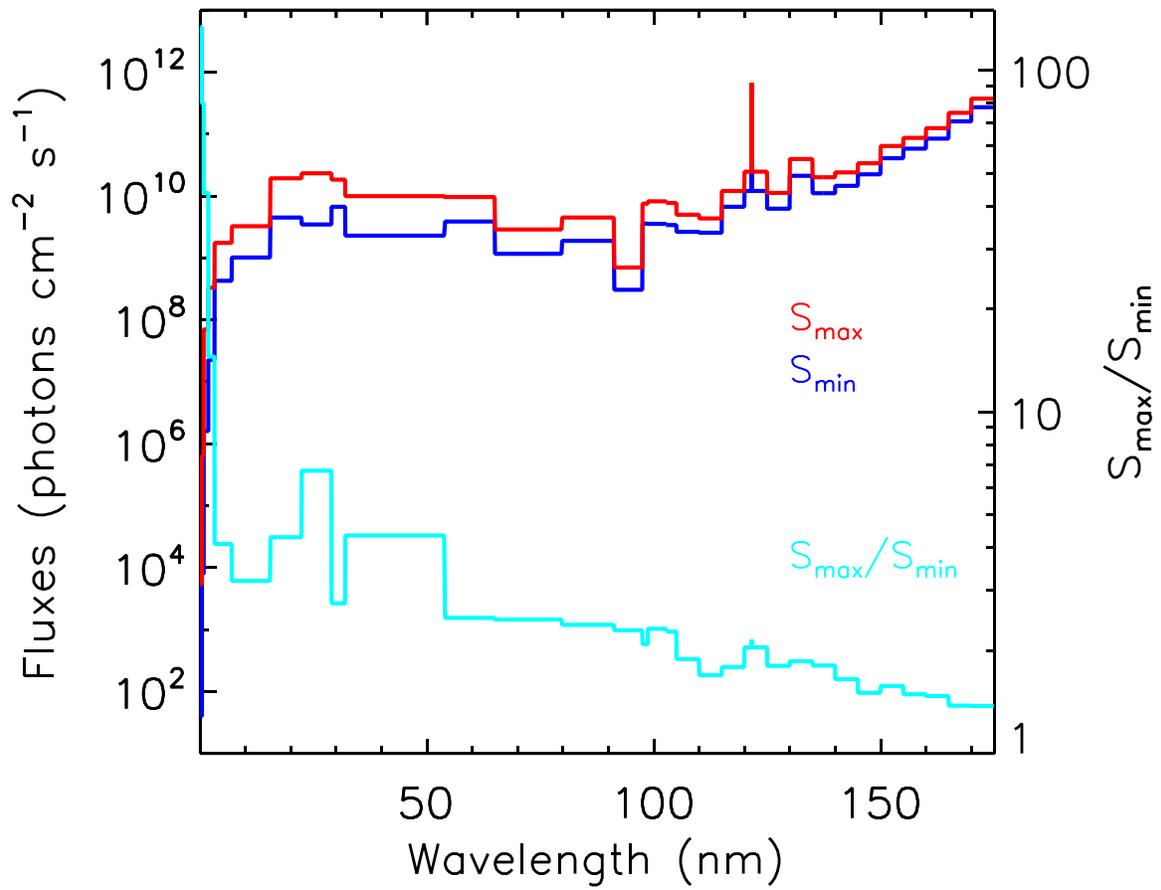

Figure 2: Solar flux inputs to the model at solar maximum (red), solar minimum (blue). Also shown is the bin-wise ratio between the fluxes at solar maximum and minimum (purple).

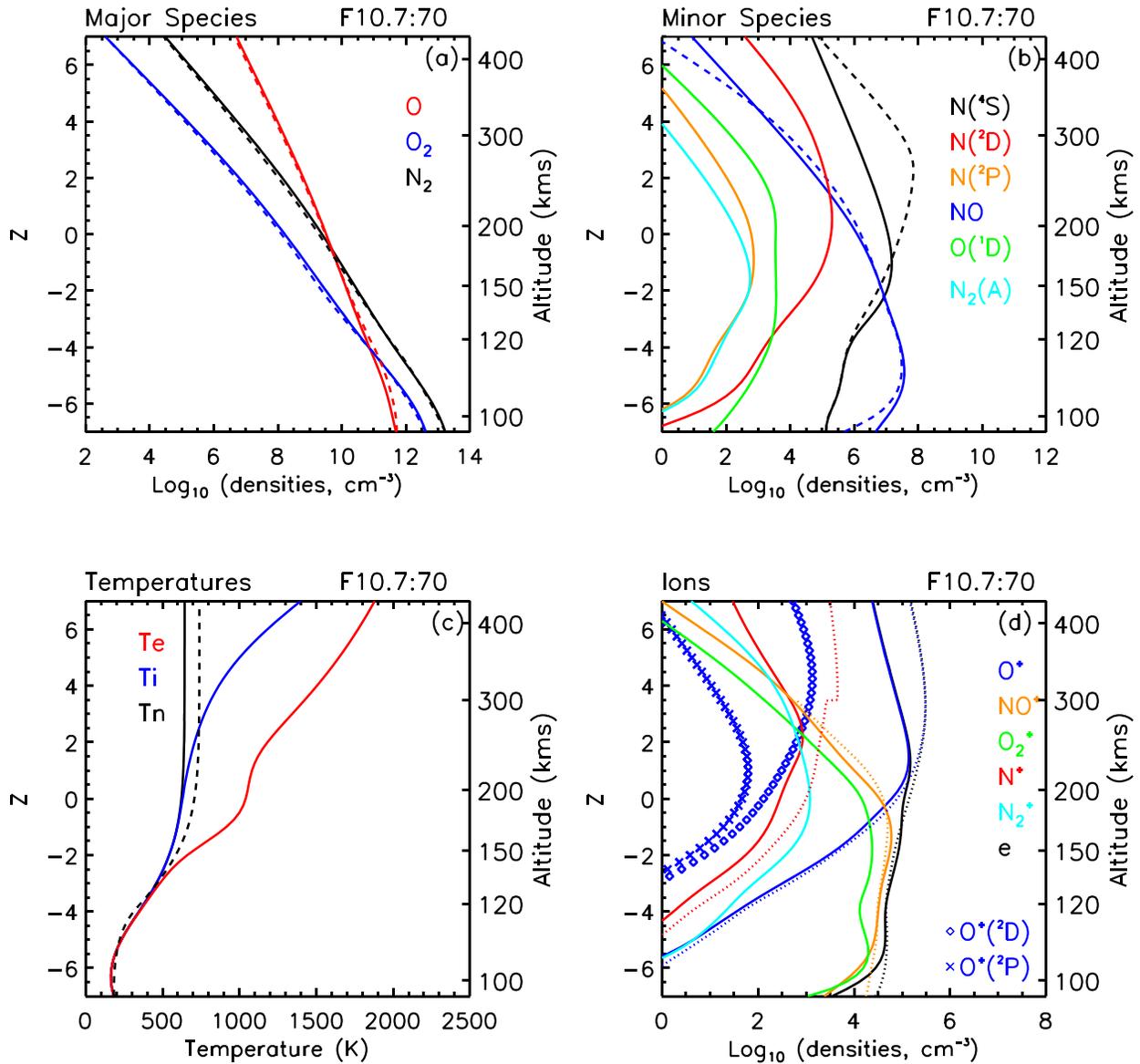

Figure 3: Overview of model outputs for solar minimum. (a) Comparisons between ACE1D calculations (solid lines) and MSIS00 results (dashed lines) of major species densities in the thermosphere (b) Calculations of minor species densities. Densities of NO and $N(^4S)$ assuming PCE are shown as dashed lines (c) Model calculations (solid line) of neutral, ion and electron temperatures. Neutral temperatures for MSIS are shown as well (dashed line) (d) Model calculations of ion densities, with dotted lines denoting densities from IRI.

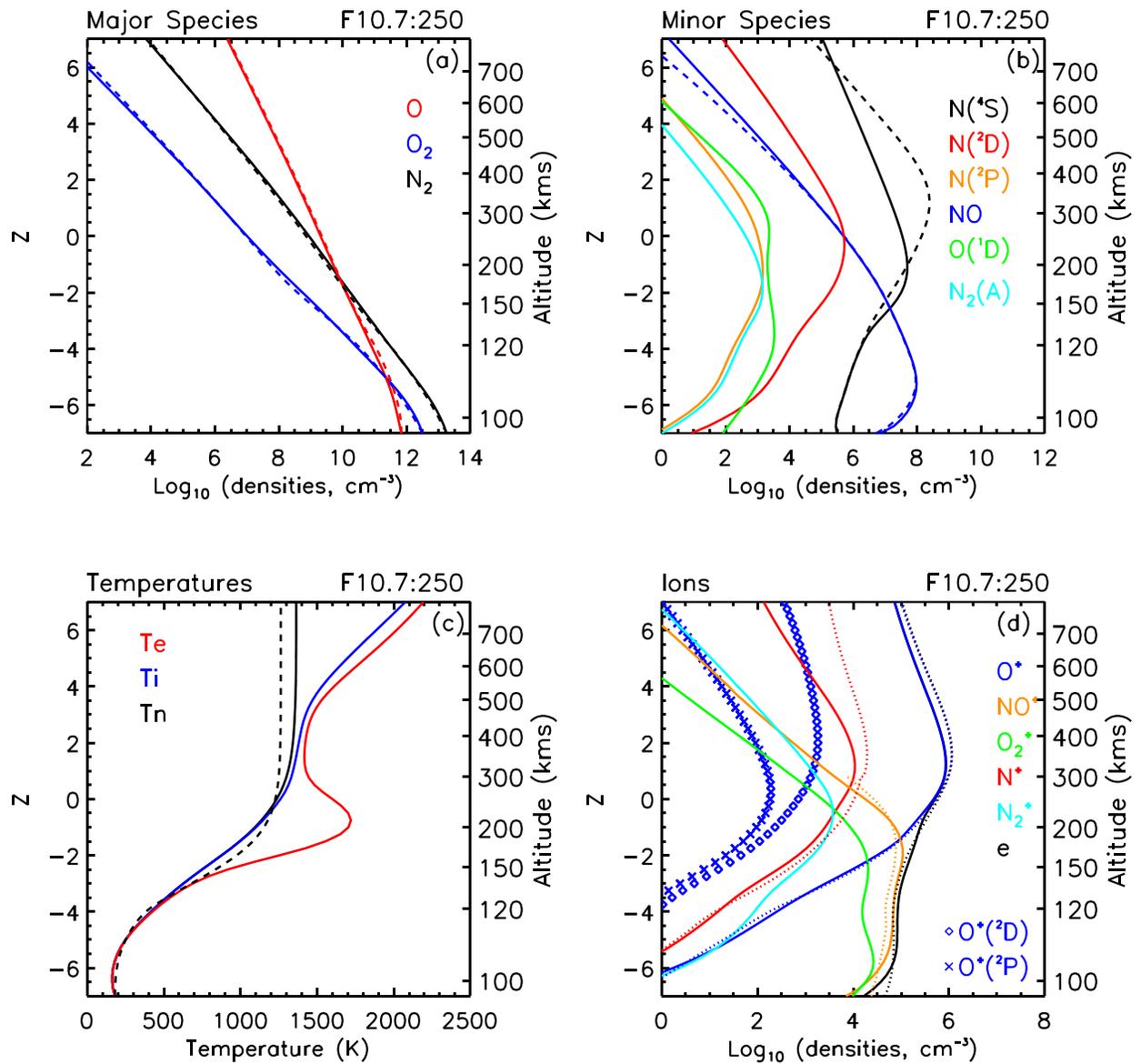

Figure 4: Overview of model outputs (same as Figure 3) for solar maximum.

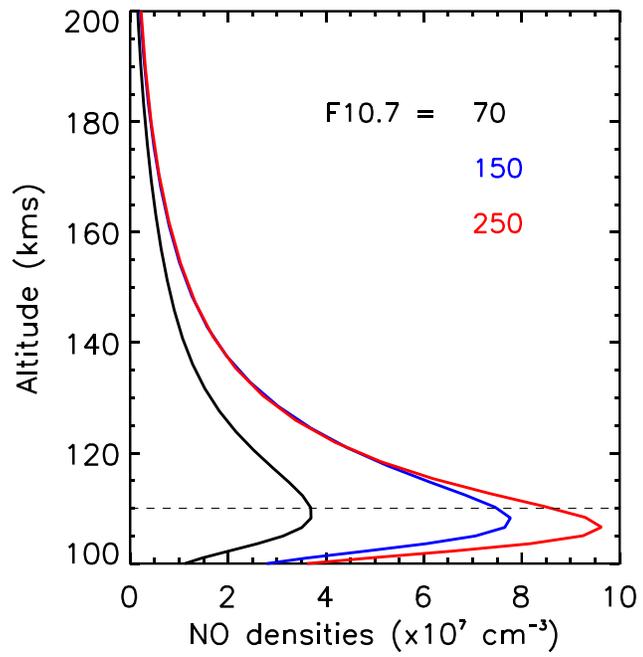

Figure 5: ACE1D model calculations of NO densities as a function of solar activity. The dashed line indicates an altitude of 110 km shown for reference.

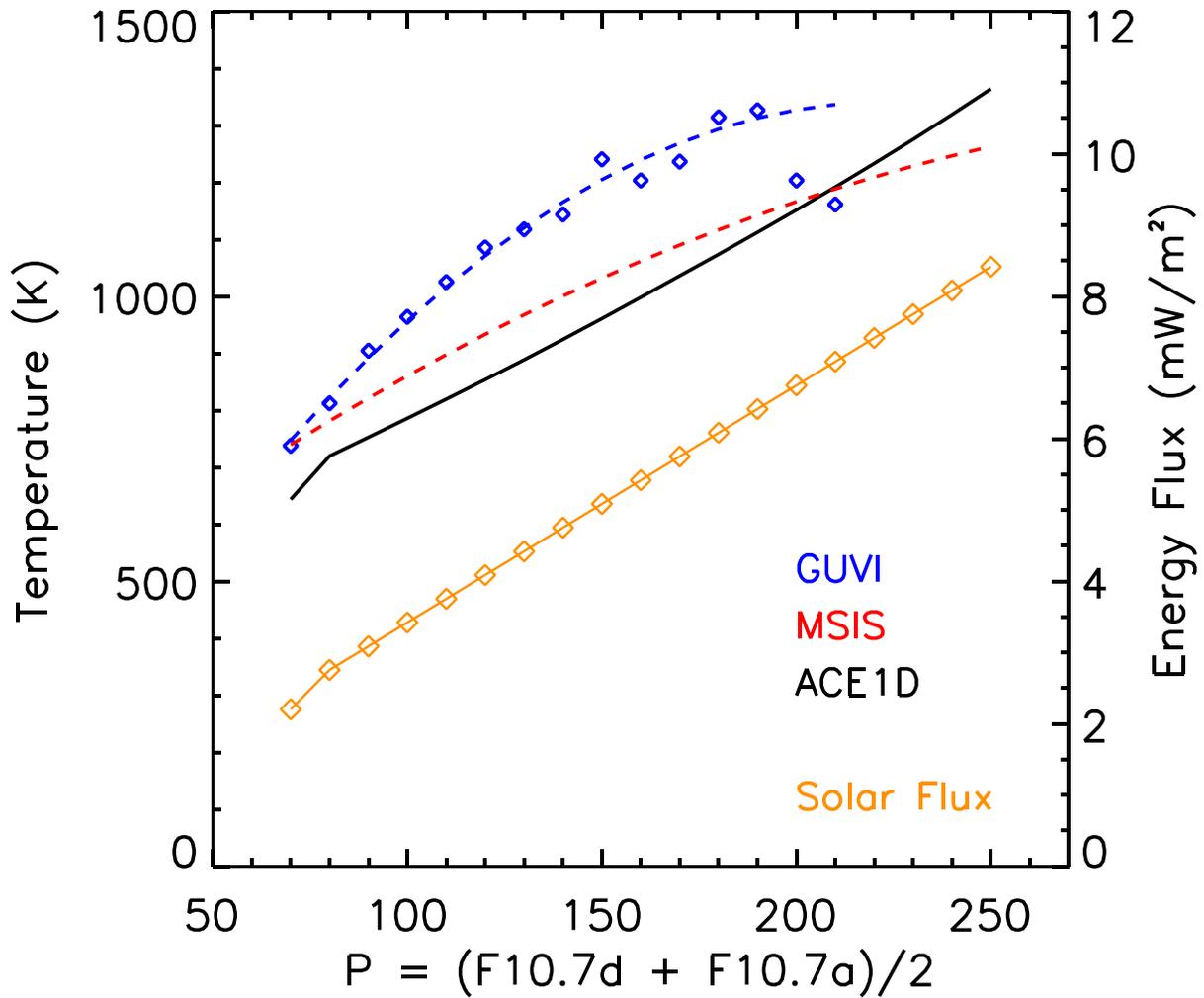

Figure 6: ACE1D model calculations (red line), MSIS data (black line) and dayside measurements by GUVI (blue diamonds) and a quadratic fit (blue dashed line) of exospheric temperatures. The fit does not use measurements for $P > 200$, which were primarily made during dawn. Also shown are integrated solar fluxes (0.05 - 110 nm) from EUVAC (orange diamonds) as a function of solar activity, which serves as the input to the ACE1D model.

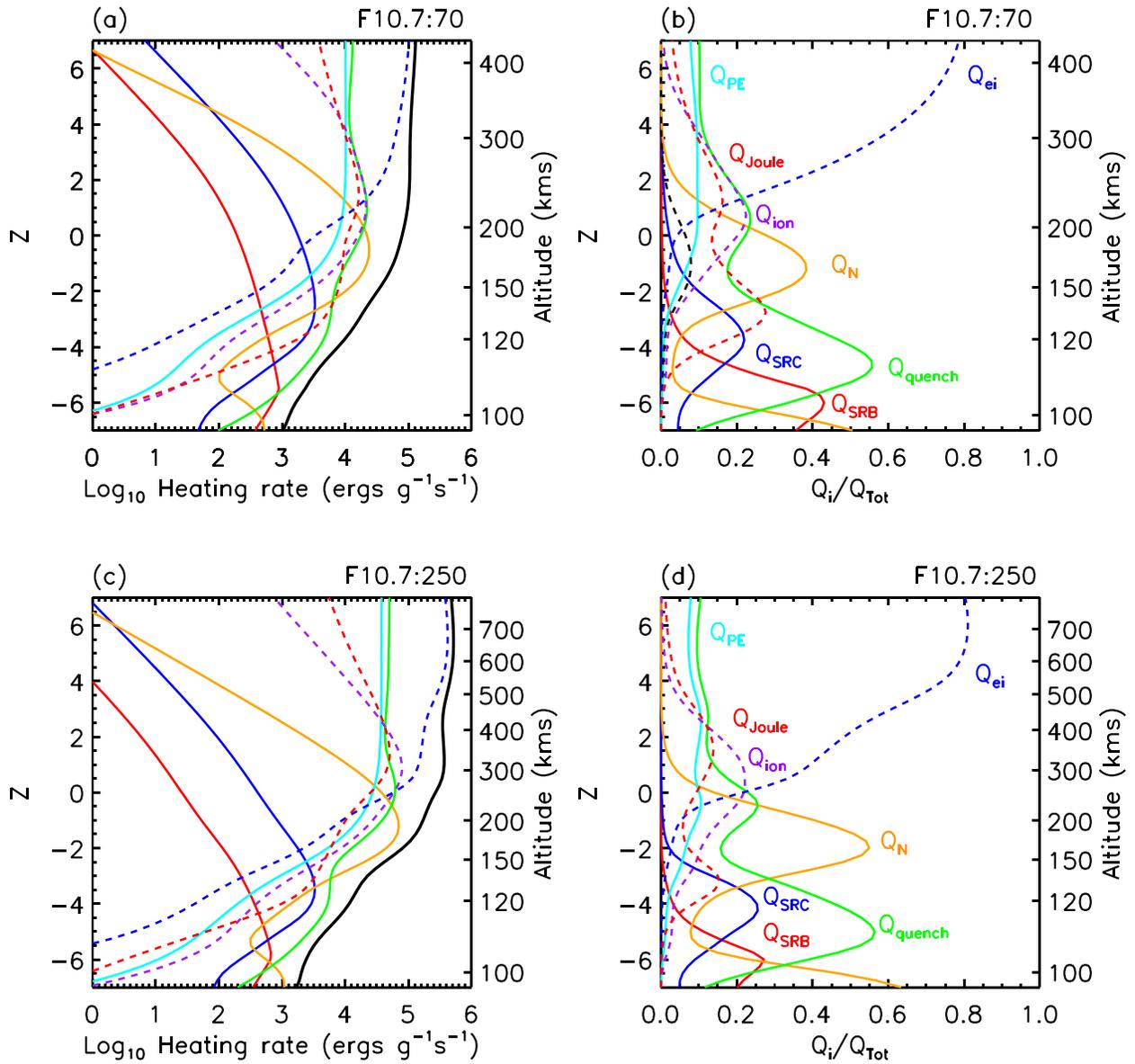

Figure 7: ACE1D model calculations of neutral gas heating rates for solar minimum (P = 70) (top) and solar maximum (P = 250) (bottom). On the left column are the magnitudes of the heating rates and on the right column are the fractional contribution of each process. Shown are the total heating rate (black), heating due to absorption in the Schumann Runge bands ($Q_{SRB}$ , red) and Schumann Runge continuum ($Q_{SRC}$ , blue), direct heating due to thermal collisions with photoelectrons ($Q_{PE}$ , aqua), exothermic reactions of neutral species ($Q_N$ , orange), quenching of excited species ($Q_{quench}$, green), exothermic ion recombination and ion-neutral reactions ($Q_i$, purple dashed), joule heating ($Q_{Joule}$, red dashed) and thermal collisions of neutrals with ions and electrons ($Q_{ei}$, blue dashed).

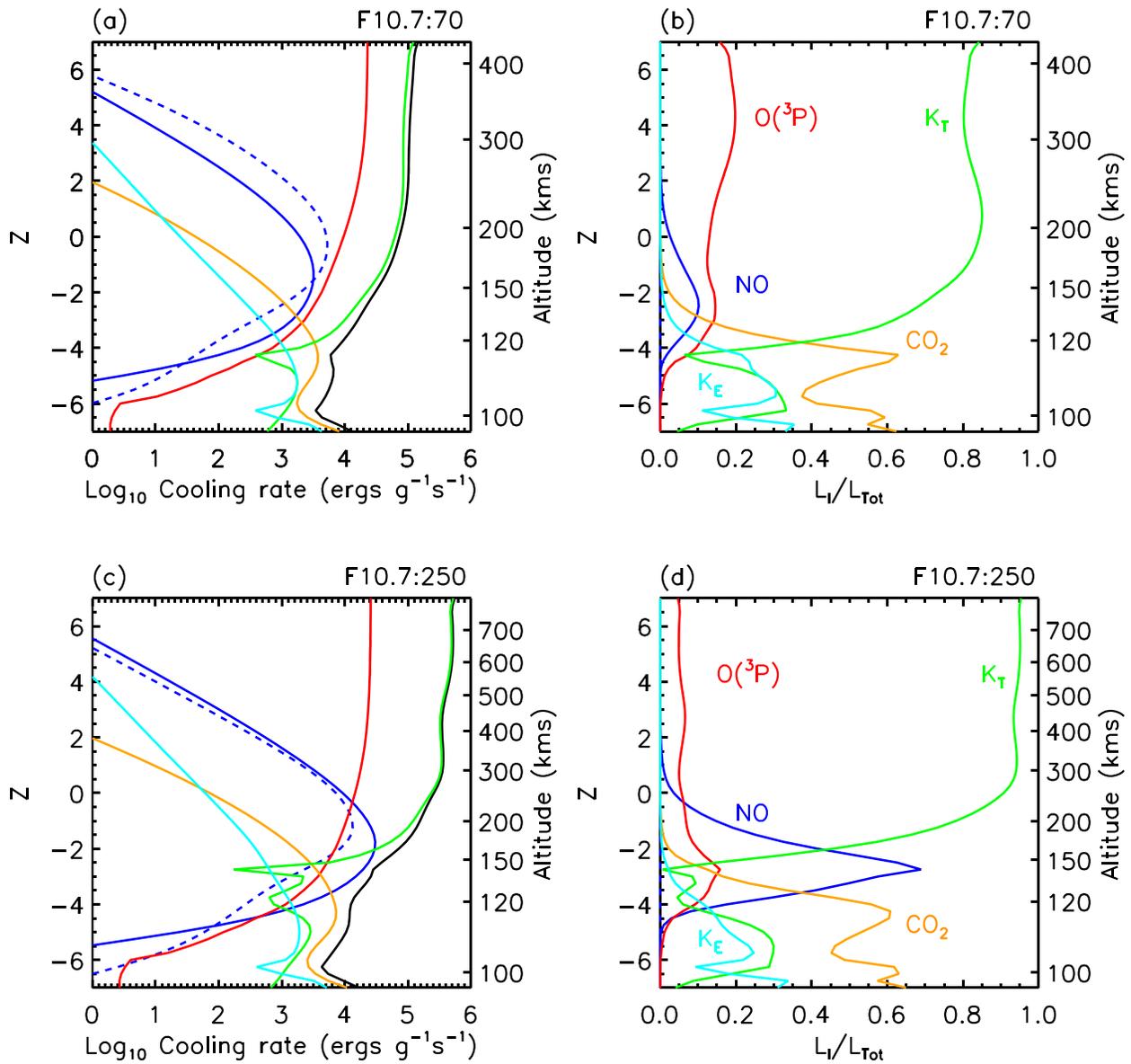

Figure 8. ACE1D model calculations of neutral gas cooling rates for solar minimum (P =

70) (top) and solar maximum (P = 250) (bottom). On the left column are the magnitudes of the cooling rates and on the right column are the fractional contribution of each process. Shown are the net cooling rate (black), heat transport due to thermal conduction ($K_T$, green) and eddy diffusion ($K_E$, aqua), and the radiative cooling due to $CO_2$ (orange), NO (blue) and $O(^1D)$ (red). NO chemiluminescence is denoted by the dashed blue line, but is not included in the calculation of the total cooling rate.

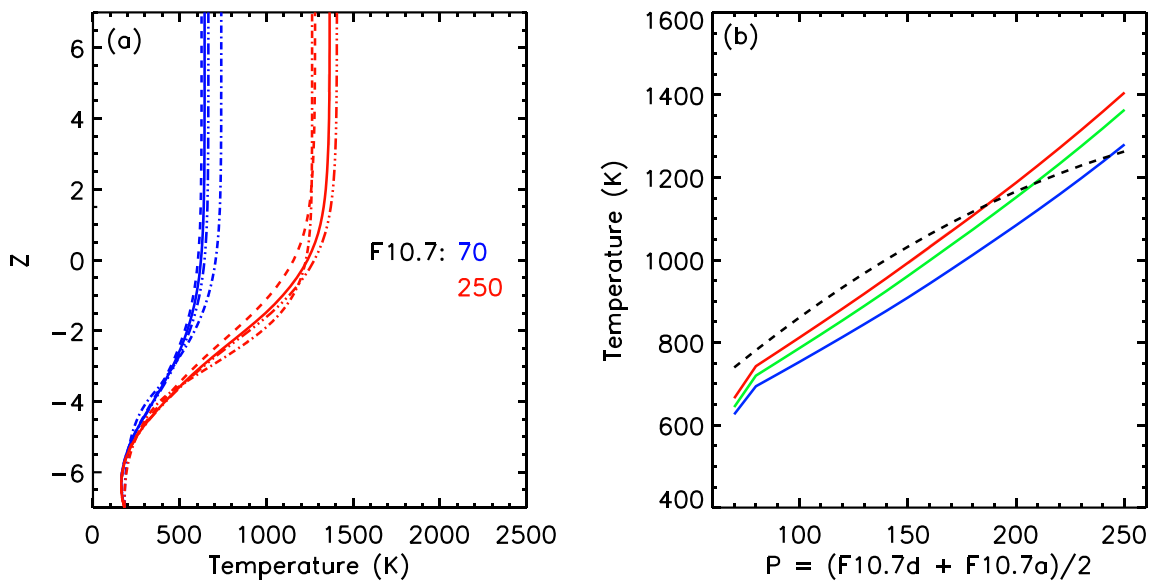

Figure 9. (a) Comparison of ACE1D neutral temperatures at solar minimum (blue) and maximum (red) for unmodified model (solid line) doubled NO cooling (dashed), excluding NO chemiluminescence (dot-dot-dash). The MSIS temperatures are shown for reference (dot-dash) (b) ACE1D calculated exospheric temperatures as a function of solar activity comparing runs with of the unmodified model (green) to doubling the NO cooling rate (blue) and excluding NO chemiluminescence effects (red). MSIS temperatures are shown for reference (dashed line).

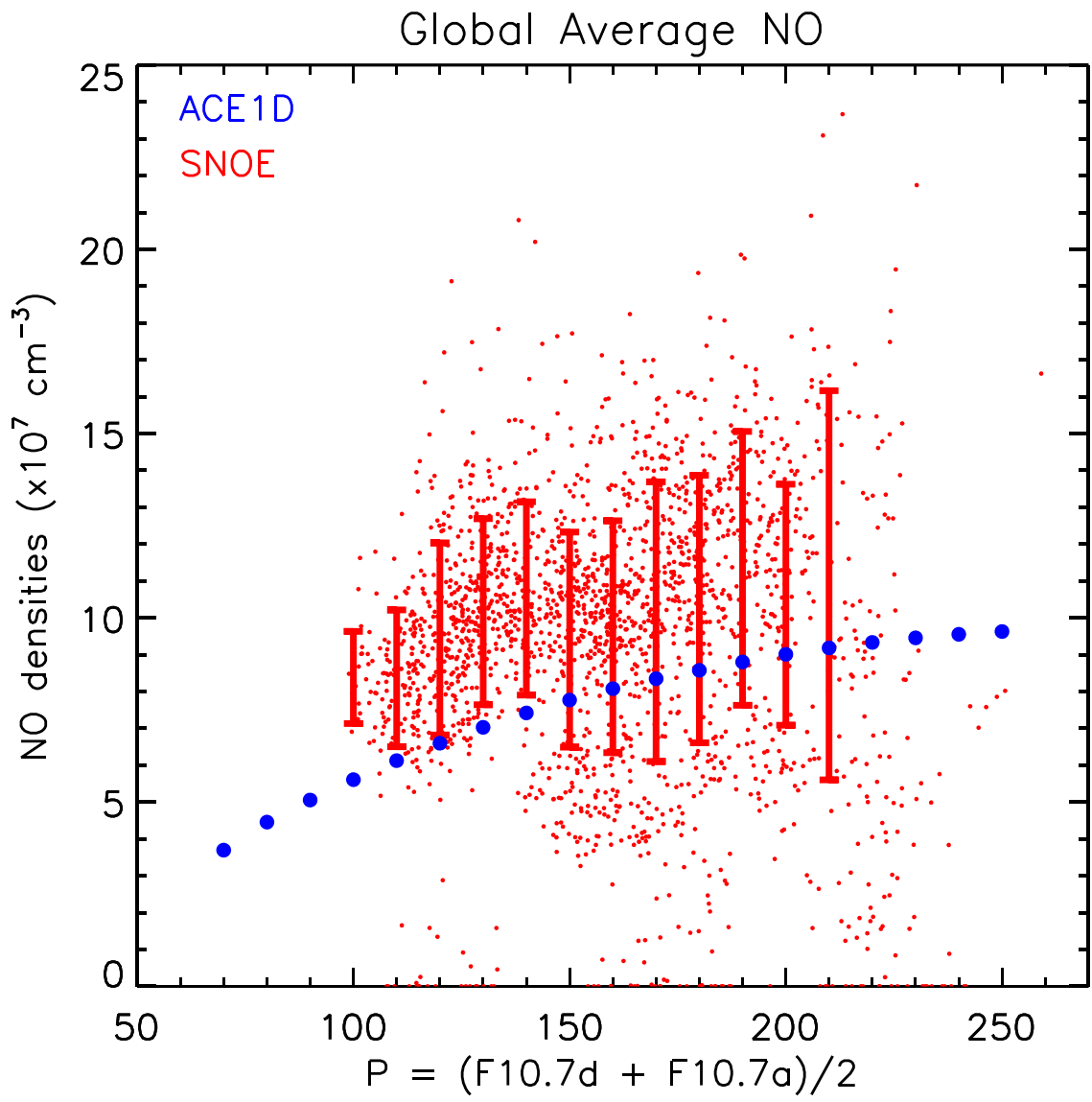

Figure 10. Comparison between measured and calculated global average peak NO densities from SNOE and ACE1D. Red points indicate measurements, along with red vertical lines that show the error bars of the measurements. Blue circles indicate modeled peak NO densities obtained from the model.

# 7 References

Bailey, Scott M., Charles A. Barth, and Stanley C. Solomon. "A model of nitric oxide in the lower thermosphere." Journal of Geophysical Research: Space Physics 107.A8 (2002): SIA-22.

Banks, P. M., and G. Kockarts. "Aeronomy, Acad." Press, New York, London (1973).

Barth, C. A., et al. "Global observations of nitric oxide in the thermosphere." Journal of Geophysical Research: Space Physics 108.A1 (2003).

Bates, D. (1951). The temperature of the upper atmosphere. Proceedings of the Physical Society. Section B, 64(9):805.

Caridade, P., Mota, V., Mohallem, J., and Varandas, A. (2008). A theoretical study of rate coefficients for the o+ no vibrational relaxation. The Journal of Physical Chemistry A, 112(5):960–965.

Castle, Karen J., et al. "Vibrational Relaxation of CO2 (v2) by O (3P) in the 142–490 K Temperature Range." *Journal of Geophysical Research: Space Physics* 117.A4 (2012).

Chamberlain, Thomas P., and Donald M. Hunten. Theory of planetary atmospheres: an introduction to their physics and chemistry. Academic Press, 1990.

Colegrove, F. D., F. S. Johnson, and W. B. Hanson. "Atmospheric composition in the lower thermosphere." Journal of Geophysical Research 71.9 (1966): 2227-2236.

DeMajistre, R., Jeng-Hwa Yee, and Xun Zhu. "Parameterizations of oxygen photolysis and energy deposition rates due to solar energy absorption in the Schumann-Runge Continuum." Geophysical research letters 28.16 (2001): 3163-3166.

Dickinson, Robert E. "Circulation and thermal structure of the Venusian thermosphere." Journal of Atmospheric Sciences 28.6 (1971): 885-894.

Dickinson, Robert E., and E. C. Ridley. "Numerical Solution for the Composition of a Thermosphere in the Presence of a Steady Subsolar to-Antisolar Circulation with Application to Venus." Journal of Atmospheric Sciences 29.8 (1972): 1557-1570.

Dickinson, Robert E., and E. C. Ridley. "A numerical model for the dynamics and composition of the Venusian thermosphere." Journal of Atmospheric Sciences 32.6 (1975): 1219-1231.

Dickinson, Robert E., E. C. Ridley, and R. G. Roble. "Meridional circulation in the thermosphere I. Equinox conditions." Journal of Atmospheric Sciences 32.9 (1975): 1737-1754.

Dickinson, Robert E., E. C. Ridley, and R. G. Roble. "A three-dimensional general circulation model of the thermosphere." Journal of Geophysical Research: Space Physics 86.A3 (1981): 1499-1512.

Dickinson, Robert E., E. C. Ridley, and R. G. Roble. "Thermospheric general circulation with coupled dynamics and composition." Journal of Atmospheric Sciences 41.2 (1984): 205-219.

Dodd, J. A., Lockwood, R. B., Hwang, E. S., Miller, S. M., & Lipson, S. J. (1999). Vibrational relaxation of NO ($\upsilon$= 1) by oxygen atoms. The Journal of chemical physics, 111(8), 3498-3507.

Duff, J. W., H. Dothe, and R. D. Sharma. "On the rate coefficient of the N (2D)+ O2→ NO+ O reaction in the terrestrial thermosphere." Geophysical research letters 30.5 (2003).

Foster, J. C., J-P. St.-Maurice, and V. J. Abreu. "Joule heating at high latitudes." Journal of Geophysical Research: Space Physics 88.A6 (1983): 4885-4897.

Grossmann, K. and Vollmann, K. (1997). Thermal infrared measurements in the middle and upper atmosphere. Advances in Space Research, 19(4):631–638.

Hedin, ALAN E. "A revised thermospheric model based on mass spectrometer and incoherent scatter data: MSIS-83." Journal of Geophysical Research: Space Physics 88.A12 (1983): 10170-10188.

Hwang, E. S., Castle, K. J., and Dodd, J. A. (2003). Vibrational relaxation of no (v= 1) by oxygen atoms between 295 and 825 k. Journal of Geophysical Research: Space Physics, 108(A3).

Huba, J. D., G. Joyce, and J. Krall. "Three-dimensional equatorial spread F modeling." Geophysical Research Letters 35.10 (2008).

Jacchia, Luigi Giuseppe. "Thermospheric temperature, density, and composition: New models." SAO special report 375 (1977).

Jacobson, M. Z. (2005). Fundamentals of atmospheric modeling. Cambridge university press.

Knipp, D. J., W. Kent Tobiska, and B. A. Emery. "Direct and indirect thermospheric heating sources for solar cycles 21–23." Solar Physics 224.1 (2004): 495-505.

Kockarts, G. "Nitric oxide cooling in the terrestrial thermosphere." Geophysical Research Letters 7.2 (1980): 137-140.

Kockarts, G. and Peetermans, W. (1970). Atomic oxygen infrared emission in the earth's upper atmosphere. Planetary and Space Science, 18(2):271–285

Lee, L. Cl, et al. "Quantum yields for the production of O (1 D) from photodissociation of O2 at 1160–1770 Å." The Journal of Chemical Physics 67.12 (1977): 5602-5606.

Liu, Han-Li, et al. "Development and validation of the Whole Atmosphere Community Climate Model with thermosphere and ionosphere extension (WACCM-X 2.0)." Journal of Advances in Modeling Earth Systems 10.2 (2018): 381-402.

Lu, G., Mlynczak, M., Hunt, L., Woods, T., and Roble, R. (2010). On the relationship of joule heating and nitric oxide radiative cooling in the thermosphere. Journal of Geophysical Research: Space Physics, 115(A5)

Meier, R. R., et al. "Remote sensing of Earth's limb by TIMED/GUVI: Retrieval of thermospheric composition and temperature." Earth and Space Science 2.1 (2015): 1-37.

Mlynczak, Marty, et al. "The natural thermostat of nitric oxide emission at 5.3 μm in the thermosphere observed during the solar storms of April 2002." Geophysical Research Letters 30.21 (2003).

Offermann, D. and Grossmann, K. (1978). Spectrometric measurement of atomic oxygen 63μm emission in the thermosphere. Geophysical Research Letters, 5(5):387–390.

Picone, J. M., et al. "NRLMSISE-00 empirical model of the atmosphere: Statistical comparisons and scientific issues." Journal of Geophysical Research: Space Physics 107.A12 (2002): SIA-15.

Qian, Liying, et al. "Calculated and observed climate change in the thermosphere, and a prediction for solar cycle 24." Geophysical research letters 33.23 (2006).

Qian, Liying, Stanley C. Solomon, and Timothy J. Kane. "Seasonal variation of thermospheric density and composition." Journal of Geophysical Research: Space Physics 114.A1 (2009).

Qian, Liying, et al. "The NCAR TIE-GCM: A community model of the coupled thermosphere/ionosphere system." Modeling the ionosphere-thermosphere system 201 (2014): 73-83.

Rawer, K., D. Bilitza, and S. Ramakrishnan. "Goals and status of the International Reference Ionosphere." Reviews of geophysics 16.2 (1978): 177-181.

Rees, M. H. and Roble, R. G. (1975). Observations and theory of the formation of stable auroral red arcs. Reviews of Geophysics, 13(1):201–242.

Richards, P. G. "Thermal electron quenching of N (2D): Consequences for the ionospheric photoelectron flux and the thermal electron temperature." Planetary and space science 34.8 (1986): 689-694.

Ridley, A. J., Y. Deng, and G. Toth. "The global ionosphere–thermosphere model." Journal of Atmospheric and Solar-Terrestrial Physics 68.8 (2006): 839-864.

Roble, R. G., E. C. Ridley, and R. E. Dickinson. "On the global mean structure of the thermosphere." Journal of Geophysical Research: Space Physics 92.A8 (1987): 8745-8758.

Roble, Raymond G. "Energetics of the mesosphere and thermosphere." The Upper Mesosphere and Lower Thermosphere: A Review of Experiment and Theory, Geophys. Monogr. Ser 87 (1995): 1-21.

Schunk, Robert W., et al. "Global assimilation of ionospheric measurements (GAIM)." Radio Science 39.1 (2004).

Schunk, Robert, and Andrew Nagy. Ionospheres: physics, plasma physics, and chemistry. Cambridge university press, 2009.

Sharma, Ramesh D., and Raymond G. Roble. "Cooling mechanisms of the planetary thermospheres: The key role of O atom vibrational excitation of $CO_2$ and NO." ChemPhysChem 3.10 (2002): 841-843.

Sharma, R., Zygelman, B., Von Esse, F., and Dalgarno, A. (1994). On the relationship between the population of the fine structure levels of the ground electronic state of atomic oxygen and the translational temperature. Geophysical research letters, 21(16):1731–1734.

Smithtro, Christopher G. Response of the ionosphere and thermosphere to extreme solar conditions. Utah State University, 2004.

Smithtro, C. G., and Jan Josef Sojka. "A new global average model of the coupled thermosphere and ionosphere." Journal of Geophysical Research: Space Physics 110.A8 (2005).

Smithtro, C. G., and S. C. Solomon. "An improved parameterization of thermal electron heating by photoelectrons, with application to an X17 flare." Journal of Geophysical Research: Space Physics 113.A8 (2008).

Solomon, S. C. and Abreu, V. J. (1989). The 630 nm dayglow. Journal of Geophysical Research: Space Physics, 94(A6):6817–6824.

Solomon, Stanley C., Paul B. Hays, and Vincent J. Abreu. "The auroral 6300 Å emission: Observations and modeling." Journal of Geophysical Research: Space Physics 93.A9 (1988): 9867-9882.

Solomon, Stanley C., and Liying Qian. "Solar extreme-ultraviolet irradiance for general circulation models." Journal of Geophysical Research: Space Physics 110.A10 (2005).

Strobel, D. F. (1978). Parameterization of the atmospheric heating rate from 15 to 120 km due to o2 and o3 absorption of solar radiation. Journal of Geophysical Research: Oceans, 83(C12):6225–6230.

Swaminathan, P. K., et al. "Model update for mesospheric/thermospheric nitric oxide." Physics and Chemistry of the Earth, Part C: Solar, Terrestrial & Planetary Science 26.7 (2001): 533-537.

Swartz, W. E. and Nisbet, J. S. (1972). Revised calculations of f region ambient electron heating by photoelectrons. Journal of Geophysical Research, 77(31):6259–6261.

Venkataramani, Karthik, Justin D. Yonker, and Scott M. Bailey. "Contribution of chemical processes to infrared emissions from nitric oxide in the thermosphere." Journal of Geophysical Research: Space Physics 121.3 (2016): 2450-2461.

Venkataramani, Karthik. Modeling the Energetics of the Upper Atmosphere. Diss. Virginia Tech, 2018.

Wise, J. O., et al. "Overview and summary of results and significant findings from the CIRRIS-1A experiment." Journal of Spacecraft and Rockets 38.3 (2001): 297-322.

Yee, Jeng-Hwa. "TIMED mission science overview." John Hopkins APL Technical Digest 24.2 (2003): 136-141.

Yonker, Justin D. Contribution of the first electronically excited state of molecular nitrogen to thermospheric nitric oxide. Diss. Virginia Polytechnic Institute and State University, 2013.

Yonker, Justin D., and Scott M. Bailey. "N2 (A) in the Terrestrial Thermosphere." Journal of Geophysical Research: Space Physics 125.1 (2020): e2019JA026508.

# A Model of the Globally-averaged Thermospheric Energy Balance

Karthik Venkataramani

Supplemental information

# Appendix A

# Model chemistry

Tabulated below are the neutral-neutral and ion-neutral reactions referred to in Figure 1, and implemented in the ACE1D model. The tables lists the rate coefficients, relaxation frequencies and exothermicities for each reaction, along with yields for various product channels, and the associated reference from which the values were obtained. $T_n$, $T_i$, and $T_e$ refer to the neutral, ion and electron temperatures respectively. The rate coefficients below are in units of $cm^3$ $s^{-1}$, except for the three body recombination reactions which are in units of $cm^6$ $s^{-1}$, and the photodissociation frequency of NO which has units of $s^{-1}$. Radiative relaxation frequencies are given in units of $s^{-1}$, and ionization cross-sections have units of $cm^{-2}$. $\beta$ denotes the branching ratio of a given product channel and is dimensionless. Where unavailable, exothermicities have been calculated using the standary enthalpy of formation of the reactants and products.

This supplemental document contains a number of references not used in the main document. The reader is directed to the Bibliography section at the end of this supplemental document for these references.

Table A.1: **Neutral-neutral reactions**

| Reaction | | | Rate coefficient | Notes | Reference |
|---|---|---|---|---|---|
| N($^2$P) + O$_2$ | $\rightarrow$ | NO+O($^3$P) | $3.1\times10^{-12}\ e^{-60/T_n}$ | | Herron (1999) |
| N($^2$P) + O($^3$P) | $\rightarrow$ | N($^4$S) + O | $2.7\times10^{-11}$ | $\beta = 0.03$ | Herron (1999) |
| | $\rightarrow$ | N($^2$D) + O | | $\beta = 0.47$ | |
| | $\rightarrow$ | NO$^+$ + e | | $\beta = 0.5$ | |
| N($^2$P) + e | $\rightarrow$ | N($^2$D) + e | $9.5\times10^{-9}\ +$ | | Berrington and Burke (1981) |
| | $\rightarrow$ | N($^4$S) + e | $1.6\times10^{-12}\ T_e^{\ 0.85}$ | | |
| N($^2$D) + N$_2$ | $\rightarrow$ | N($^4$S) + N$_2$ | $1.74\times10^{-14}$ | | Herron (1999) |
| N($^2$D) + O$_2$ | $\rightarrow$ | NO(v) + O($^3$P) | $6.2\times10^{-12}(T_n/300)$ | | Duff et al. (2003) |
| N($^2$D) + O | $\rightarrow$ | N($^4$S) + O | $1.65\times10^{-12}\ e^{-260/T_n}$ | | Fell et al. (1990) |
| N($^2$D) + NO | $\rightarrow$ | N$_2$ + O($^3$P) | $6.7\times10^{-11}$ | | Herron (1999) |
| N($^2$D) + e | $\rightarrow$ | N($^4$S) + e | $3.86\times10^{-10}\ (T_e/300)^{0.81}$ | | Berrington and Burke (1981) |
| N($^4$S) + O$_2$ | $\rightarrow$ | NO(v) + O($^3$P) | $1.5\times10^{-11}\ e^{-3600/T_n}$ | | Sander et al. (2006) |
| N($^4$S) + NO | $\rightarrow$ | N$_2$ + O($^3$P) | $2.1\times10^{-11}\ e^{100/T_n}$ | | Sander et al. (2006) |
| O + O + M | $\rightarrow$ | O$_2$ + M | $9.59\times10^{-34}\ e^{480/T_n}$ | | Logan et al. (1978) |
| O + O$_2$ + M | $\rightarrow$ | O$_3$ + M | $6\times10^{-34}\ (T_n/300)^{-2.4}$ | | Sander et al. (2006) |
| N$_2$(A) + O | $\rightarrow$ | NO + N($^2$D) | $1\times10^{-13}$ | | Dilecce and De Benedictis (1999) |
| N($^2$P) + N$_2$ | $\rightarrow$ | N($^4$S) + N$_2$ | $5\times10^{-17}$ | | Herron (1999) |
| N($^2$P) + NO | $\rightarrow$ | N($^4$S) + NO | $2.9\times10^{-11}$ | | Herron (1999) |

A.1 – continued from previous page

| Reaction | | | Rate coefficient | Notes | Reference |
|---|---|---|---|---|---|
| $O(^1D) + N_2$ | $\rightarrow$ | $O(^3P) + N_2$ | $2 \times 10^{-11} \, e^{107.8/T_n}$ | | Fennelly et al. (1994) |
| $O(^1D) + O_2$ | $\rightarrow$ | $O(^3P) + O_2$ | $2.9 \times 10^{-11} \, e^{67.5/T_n}$ | | Fennelly et al. (1994) |
| $O(^1D) + O$ | $\rightarrow$ | $O(^3P) + O$ | $8 \times 10^{-12}$ | | Fennelly et al. (1994) |
| $N_2 + h\nu / e^*$ | $\rightarrow$ | $2N(^2P, \, ^2D, \, ^4S)$ | | $\beta_{N(^2P)} = 0.224$ | Yonker (2013) |
| | | | | $\beta_{N(^2D)} = 0.276$ | |
| | | | | $\beta_{N(^4S)} = 0.5$ | |
| | $\rightarrow$ | $N_2^+ + e$ | | | Solomon and Qian (2005) |
| | $\rightarrow$ | $N + N^+ + e$ | | | |
| $O_2 + h\nu / e^*$ | $\rightarrow$ | $O(^1D) + O(^3P)$ | | | Lee et al. (1977), |
| | $\rightarrow$ | $O_2^+ + e$ | | | Solomon and Qian (2005) |
| | $\rightarrow$ | $O + O^+ + e$ | | | |
| $O + h\nu / e^*$ | $\rightarrow$ | $O^+ + e$ | | | Solomon and Qian (2005) |
| $N + h\nu / e^*$ | $\rightarrow$ | $N^+ + e$ | | | Solomon and Qian (2005) |
| $NO + h\nu / e^*$ | $\rightarrow$ | $NO^+ + e$ | $J_{NO} = 4.5 \times 10^{-6}$ | $\sigma_i = 2.02 \times 10^{-18}$ | Watanabe (1958) |
| | $\rightarrow$ | $N(^4S) + O$ | | | Barth (1992) |

Table A.2: **Ion-neutral reactions**

| Reaction | | Rate coefficient | Notes | Reference |
|---|---|---|---|---|
| $N^+ + NO$ | $\rightarrow$ $N(^4S) + NO^+$ | $6.5\times10^{-9}(T_i)^{-0.44}$ | $\beta = 0.91$ | Midey et al. (2004) |
| | $\rightarrow$ $O + N_2^+$ | | $\beta = 0.07$ | |
| | $\rightarrow$ $O^+ + N_2$ | | $\beta = 0.02$ | |
| $N^+ + O(^3P)$ | $\rightarrow$ $N(^4S) + O^+$ | $4.5\times10^{-12}$ | | Anicich (2003) |
| $N^+ + O_2$ | $\rightarrow$ $N(^2D) + O_2^+$ | $5.5\times10^{-10}$ | $\beta = 0.5$ | Midey et al. (2006) |
| | $\rightarrow$ $NO^+ + O$ | | $\beta = 0.42$ | |
| | $\rightarrow$ $NO + O^+(^4S)$ | | $\beta = 0.08$ | |
| $N_2^+ + e$ | $\rightarrow$ $N(^2D) + N(^2D)$ | $2.2\times10^{-7}(300/T_e)^{0.39}$ | $\beta = 0.52$ | Sheehan and St-Maurice (2004), |
| | $\rightarrow$ $N(^2D) + N(^4S)$ | | $\beta = 0.37$ | Peterson et al. (1998) |
| | $\rightarrow$ $N(^2P) + N(^4S)$ | | $\beta = 0.11$ | |
| $N_2^+ + N(^4S)$ | $\rightarrow$ $N^+ + N(^4S)$ | $1\times10^{-11}$ | | Fox and Sung (2001) |
| $N_2^+ + NO$ | $\rightarrow$ $NO^+ + N_2$ | $7.5\times10^{-9}(T_i)^{-0.52}$ | | Midey et al. (2004) |
| $N_2^+ + O(^3P)$ | $\rightarrow$ $NO^+ + N(^2D)$ | $1.33\times10^{-10}(300/T_i)^{0.44}$ | $T_i \leq 1500$ K | Fox and Sung (2001) |
| | $\rightarrow$ $NO^+ + N(^4S)$ | $6.5\times10^{-11}(T_i/300)^{0.2}$ | $T_i > 1500$ K | |
| | $\rightarrow$ $NO^+ + N(^2D)$ | | $\beta = 0.9$ | |
| | $\rightarrow$ $NO^+ + N(^4S)$ | | $\beta = 0.05$ | |
| | $\rightarrow$ $O^+ + N_2$ | | $\beta = 0.05$ | |

A.2 – continued from previous page

| Reaction | | | Rate coefficient | Notes | Reference |
|---|---|---|---|---|---|
| $N_2^+ + O_2$ | $\rightarrow$ | $N_2 + O_2^+$ | $5.1\times10^{-11}(300/T_i)^{1.16}$ | $T_i \leq 1000$ K | Fox and Sung (2001) |
| | | | $1.26\times10^{-11}(T_i/1000)^{0.67}$ | $T_i > 1000$ K | |
| $NO^+ + e$ | $\rightarrow$ | $N(^2D) + O$ | $3.5\times10^{-7}(300/T_e)^{0.69}$ | $\beta = 0.95$ | Sheehan and St-Maurice (2004) |
| | $\rightarrow$ | $N(^4S) + O$ | | $\beta = 0.05$ | Hellberg et al. (2003) |
| $O^+ + N(^2D)$ | $\rightarrow$ | $N^+ + O$ | $1.3\times10^{-10}$ | | Richards and Voglozin (2011) |
| $O^+ + N_2$ | $\rightarrow$ | $NO^+ + N(^4S)$ | $1\times10^{-12}\,(300/T_i)^{0.45}$ | $T_i \leq 1000$ | Richards and Voglozin (2011) |
| | | | $7\times10^{-13}\,(T_i/1000)^{2.12}$ | $T_i > 1000$ | Richards and Voglozin (2011) |
| $O^+ + NO$ | $\rightarrow$ | $NO^+ + O$ | $5.01\times10^{-13}\,(300/T_i)^{1.68}$ | | Dotan and Viggiano (1999) |
| $O^+ + O_2$ | $\rightarrow$ | $O_2^+ + O$ | $1.7\times10^{-11}\,(300/T_i)^{0.77} + 8.54\times10^{-11}\,e^{-3461/T_i}$ | | Hierl et al. (1997) |
| $O^+(^2D) + e$ | $\rightarrow$ | $O^+(^4S) + e$ | $6.03\times10^{-8}(300/T_e)^{0.5}$ | | Richards and Voglozin (2011) |
| $O^+(^2D) + N(^4S)$ | $\rightarrow$ | $N^+ + O$ | $1.5\times10^{-10}$ | | Richards and Voglozin (2011) |
| $O^+(^2D) + N_2$ | $\rightarrow$ | $O(^3P) + N_2^+$ | $5.7\times10^{-10}\,e^{-400/T_i}$ | | Fox and Sung (2001) |
| $O^+(^2D) + NO$ | $\rightarrow$ | $NO^+ + O$ | $1.2\times10^{-9}$ | | Richards and Voglozin (2011) |
| $O^+(^2D) + O(^3P)$ | $\rightarrow$ | $O^+(^4S) + O(^3P)$ | $1\times10^{-11}$ | | Richards and Voglozin (2011) |
| $O^+(^2D) + O_2$ | $\rightarrow$ | $O_2^+ + O$ | $7\times10^{-10}$ | | Johnsen and Biondi (1980) |
| $O^+(^2P) + e$ | $\rightarrow$ | $O^+(^4S) + e$ | $1.84\times10^{-7}(300/T_e)^{0.5} +$ | | Richards and Voglozin (2011) |
| | $\rightarrow$ | $O^+(^2D) + e$ | $3.03\times10^{-8}(300/T_e)^{0.5}$ | | |

A.2 – continued from previous page

| Reaction | | Rate coefficient | Notes | Reference |
|---|---|---|---|---|
| $O^+(^2P) + N_2$ | $\rightarrow$ $O(^3P) + N_2^+$ | $5.7\times10^{-10}\, e^{-400/T_i}$ | | Fox and Sung (2001) |
| $O^+(^2P) + O(^3P)$ | $\rightarrow$ $O^+(^4S) + O(^3P)$ | $5\times10^{-11}$ | | Stephan et al. (2003) |
| $O^+(^2P) + O_2$ | $\rightarrow$ $O_2^+ + O$ | $7\times10^{-10}$ | | Johnsen and Biondi (1980) |
| $O_2^+ + e$ | $\rightarrow$ | $2\times10^{-7}(300/T_e)^{0.7}$ | $T_e < 1200$ K | Fennelly et al. (1994) |
| | $\rightarrow$ | $1.6\times10^{-7}(300/T_e)^{0.55}$ | $T_e \geq 1200$ K | |
| | $\rightarrow$ $O(^3P)+O(^3P)$ | $\beta = 0.22$ | | Schunk and Nagy (2009) |
| | $\rightarrow$ $O(^3P)+O(^1D)$ | $\beta = 0.42$ | | |
| | $\rightarrow$ $O(^1D)+O(^1D)$ | $\beta = 0.36$ | | |
| $O_2^+ + N(^2D)$ | $\rightarrow$ $NO^+ + O$ | $1.8\times10^{-10}$ | | Richards and Voglozin (2011) |
| $O_2^+ + N(^2P)$ | $\rightarrow$ $O_2^+ + N$ | $2.2\times10^{-11}$ | | Zipf et al. (1980) |
| $O_2^+ + N(^4S)$ | $\rightarrow$ $NO^+ + O$ | $1.33\times10^{-10}$ | | Scott et al. (1998) |
| $O_2^+ + NO$ | $\rightarrow$ $NO^+ + O_2$ | $4.5\times10^{-10}$ | | Midey and Viggiano (1999) |

Table A.3: **Loss frequencies and Photon Energies**

| Reaction | | Loss frequency ($s^{-1}$) | $h\nu$ Energy (eV) | Reference |
|---|---|---|---|---|
| $N(^2D)$ | $\rightarrow$ $N(^4S) + h\nu$ | $1.279\times10^{-5}$ | 2.38 | Swaminathan et al. (1998) |
| $N(^2P)$ | $\rightarrow$ $N(^4S) + h\nu$ | $5.4\times10^{-3}$ | 3.57 | Swaminathan et al. (1998) |

A.3 – continued from previous page

| Reaction | | Relaxation frequency (s⁻¹) | $h\nu$ Energy (eV) | Reference |
|---|---|---|---|---|
| | $\rightarrow$ N($^2$D) + $h\nu$ | $8.054 \times 10^{-2}$ | 1.19 | Roble (1995) |
| O$^+$($^2$P) $\rightarrow$ | O$^+$($^4$S) + $h\nu$ | 0.047 | 5.02 | |
| | $\rightarrow$ O$^+$($^2$D) + $h\nu$ | 0.171 | 1.69 | |
| O($^1$D) $\rightarrow$ | O($^3$P) + $h\nu$ | 0.0059 | 1.97 | Fennelly et al. (1994) |

Table A.4: **Reaction Exothermicities**

| Reaction | | Exothermicity (eV) | Reference |
|---|---|---|---|
| N$^+$ + NO $\rightarrow$ | N($^4$S) + NO$^+$ | 5.3 | Midey et al. (2004) |
| | $\rightarrow$ O + N$_2^+$ | 2.2 | |
| | $\rightarrow$ O$^+$ + N$_2$ | 4.2 | |
| N$^+$ + O $\rightarrow$ | O$^+$($^4$S) + N($^4$S) | 0.98 | Roble (1995) |
| N$^+$ + O$_2$ $\rightarrow$ | N($^2$D) + O$_2^+$ | 0.04 | Midey et al. (2006) |
| | $\rightarrow$ NO$^+$ + O | 6.7 | |
| | $\rightarrow$ NO + O$^+$($^4$S) | 2.3 | |
| N$_2^+$ + e $\rightarrow$ | N($^2$D) + N($^2$D) | 3.44 | Roble (1995) |
| | $\rightarrow$ N($^2$D) + N($^4$S) | 5.82 | |
| | $\rightarrow$ N($^2$P) + N($^4$S) | 4.63 | |

A.4 – continued from previous page

| Reaction | | | Exothermicity (eV) | Reference |
|---|---|---|---|---|
| $N_2^+ + NO$ | $\rightarrow$ | $NO^+ + N_2$ | 6.3 | Midey et al. (2004) |
| $N_2^+ + O$ | $\rightarrow$ | $NO^+ + N(^2D)$ | 0.646 | Scott et al. (1999) |
| | $\rightarrow$ | $NO^+ + N(^4S)$ | 3.06 | |
| | $\rightarrow$ | $O^+ + N_2$ | 1.02 | |
| $N_2^+ + O_2$ | $\rightarrow$ | $O_2^+ + N_2$ | 3.5 | Dotan et al. (1997) |
| $NO^+ + e$ | $\rightarrow$ | $N(^2D) + O$ | 0.38 | Roble (1995) |
| | $\rightarrow$ | $N(^4S) + O$ | 2.75 | |
| $O^+ + N(^2D)$ | $\rightarrow$ | $N^+ + O$ | 1.45 | Roble (1995) |
| $O^+ + N_2$ | $\rightarrow$ | $NO^+ + N(^4S)$ | 1.09 | Roble (1995) |
| $O^+ + NO$ | $\rightarrow$ | $NO^+ + O$ | 4.3 | Dotan and Viggiano (1999) |
| $O^+ + O_2$ | $\rightarrow$ | $O_2^+ + O$ | 1.56 | Roble (1995) |
| $O^+(^2D) + e$ | $\rightarrow$ | $O^+(^4S) + e$ | 3.31 | Roble (1995) |
| $O^+(^2D) + N_2$ | $\rightarrow$ | $O(^3P) + N_2^+$ | 1.35 | Roble (1995) |
| $O^+(^2D) + O(^3P)$ | $\rightarrow$ | $O^+(^4S) + O(^3P)$ | 3.31 | Stephan et al. (2003) |
| $O^+(^2D) + O_2$ | $\rightarrow$ | $O_2^+ + O$ | 4.865 | Roble (1995) |
| $O^+(^2P) + e$ | $\rightarrow$ | $O^+(^4S) + e$ | 5.0 | Stephan et al. (2003) |
| | $\rightarrow$ | $O^+(^2D) + e$ | 1.69 | |
| $O^+(^2P) + O(^3P)$ | $\rightarrow$ | $O^+(^4S) + O(^3P)$ | 5.0 | Roble (1995) |

A.4 – continued from previous page

| Reaction | | | Exothermicity (eV) | Reference |
|---|---|---|---|---|
| $O^+(^2P) + N_2$ | $\rightarrow$ | $O(^3P) + N_2^+$ | 3.05 | Li et al. (1997) |
| $O_2^+ + e$ | $\rightarrow$ | $O(^3P) + O(^3P)$ | 6.99 | Schunk and Nagy (2009) |
| | $\rightarrow$ | $O(^3P) + O(^1D)$ | 5.02 | |
| | $\rightarrow$ | $O(^1D) + O(^1D)$ | 3.06 | |
| $O_2^+ + N(^2P)$ | $\rightarrow$ | $O_2^+ + N$ | 3.57 | Zipf et al. (1980) |
| $O_2^+ + N(^4S)$ | $\rightarrow$ | $NO^+ + O$ | 4.18 | Scott et al. (1998) |
| $O_2^+ + NO$ | $\rightarrow$ | $NO^+ + O_2$ | 2.81 | Midey and Viggiano (1999) |
| $N(^4S) + O_2$ | $\rightarrow$ | $NO(v) + O(^3P)$ | 1.385 | Kennealy et al. (1978) |
| $N(^4S) + NO$ | $\rightarrow$ | $N_2 + O(^3P)$ | 3.25 | Baulch et al. (2005) |
| $N(^2D) + O_2$ | $\rightarrow$ | $NO(v) + O(^3P)$ | 3.76 | Kennealy et al. (1978) |
| $N(^2D) + NO$ | $\rightarrow$ | $N_2 + O(^3P)$ | 5.63 | Roble (1995) |
| $N(^2P) + O_2$ | $\rightarrow$ | $NO+O(^3P)$ | 4.96 | Rawlins et al. (1989) |
| $O + O + M$ | $\rightarrow$ | $O_2 + M$ | 5.12 | Roble (1995) |
| $N(^2D) + M$ | $\rightarrow$ | $N(^4S) + M$ | 2.38 | Roble (1995), Herron (1999) |
| $(M = O, N_2, e)$ | | | | |
| $N(^2P) + N_2$ | $\rightarrow$ | $N(^4S) + N_2$ | 3.57 | Herron (1999), Zipf et al. (1980) |
| $O(^1D) + M$ | $\rightarrow$ | $O(^3P) + M$ | 1.97 | Roble (1995) |

# Appendix B

# Ionospheric parameters

The parameters concerning the heating of ionospheric electrons, joule heating of neutral species, heating and cooling terms associated with thermal collisions involving neutrals, electrons and ions, and the diffusion of ions are listed here.

## B.1   Photoelectron heating rate of ionospheric electrons

The ACE1D model uses the parameterization provided by Smithtro and Solomon (2008) to obtain the electron heating rate due to energetic photoelectrons. In this method, the heating rate is calculated by scaling the photoionization rate due to solar photons between 0-55 and 55-105 nm:

$$Q_{0-55nm} = <\hat{\epsilon}> \sum_i P_i \bar{\epsilon}_i \tag{B.1}$$

$$Q_{55-105nm} = <\hat{\epsilon}> P_i \tag{B.2}$$

$$\log_e <\hat{\epsilon}> = \sum_i c_i (\log_e R)^i \tag{B.3}$$

$$R = \frac{n_e}{[N_2] + [O_2] + [O]} \tag{B.4}$$

Where $Q$ denotes the heating rate in units of eV cm$^3$ s$^{-1}$, $R$ is the deposition parameter, $<\hat{\epsilon}>$ is the heating efficiency factor and $\bar{\epsilon}_i$ is the photon energy associated with the bin that produces a local photoionization rate of $P_i$. The summation for $Q_{0-55nm}$ is over all wavelength bins. The coefficients for the polynomial fit $c_i$ are given in Table B.1.

Table B.1: Fit parameters for photoelectron heating

|        | 0 - 55 nm              | 55 - 105 nm            |
|--------|------------------------|------------------------|
| c0     | 1.468                  | 1.020                  |
| c1     | $9.229\times10^{-1}$   | $1.540\times10^{-2}$   |
| c2     | $4.956\times10^{-2}$   | $-6.858\times10^{-3}$  |
| c3     | $-1.897\times10^{-2}$  | $-8.528\times10^{-3}$  |
| c4     | $-3.934\times10^{-3}$  | $-2.052\times10^{-3}$  |
| c5     | $-2.643\times10^{-4}$  | $-1.634\times10^{-4}$  |
| c6     | $-5.980\times10^{-6}$  | $-4.314\times10^{-6}$  |

The total electron volume heating rate is simply the sum of the individual heating rates:

$$Q = Q_{0-55\text{nm}} + Q_{55-105\text{nm}} \tag{B.5}$$

## B.2 Electron cooling rates

Elastic and inelastic collisions of electrons with background neutral species is the dominant energy loss mechanism for electrons in the thermosphere, expressions for which are given here in units of eV cm$^{-3}$ s$^{-1}$. $n_e$ and n(M) in the below expressions refer to the electron density and the density of species M respectively.

### B.2.1 $N_2$ rotation

$$L_e(N_2) = 3.5 \times 10^{-14} \ n_e n(N_2)(T_e - T_n)/T_e^{0.5} \tag{B.6}$$

### B.2.2 $O_2$ rotation

$$L_e(O_2) = 5.2 \times 10^{-15} \ n_e n(O_2)(T_e - T_n)/T_e^{0.5} \tag{B.7}$$

### B.2.3 $CO_2$ rotation

$$L_e(CO_2) = 5.8 \times 10^{-14} \ n_e n(CO_2)(T_e - T_n)/T_e^{0.5} \tag{B.8}$$

## B.2.4 $N_2$ vibration

$$L_e(N_2) = n_e n(N_2)[1 - \exp(-E_1/T_{vib})] \times \sum_{v=1}^{10} Q_{0v}[1 - \exp[vE_1(T_e^{-1} - T_{vib}^{-1})]]$$

$$+ n_e n(N_2)[1 - \exp(-E_1/T_{vib})] \exp(-E_1/T_{vib)} \tag{B.9}$$

$$\times \sum_{v=2}^{9} Q_{1v}[1 - \exp[(v-1)E_1(T_e^{-1} - T_{vib}^{-1})]]$$

where $E_1 = 3353$ K (0.2889 eV), $T_{vib} = T_n$, and

$$\log Q_{0v} = A_{0v} + B_{0v}T_e + C_{0v}T_e^2 + D_{0v}T_e^3 + F_{0v}T_e^4 - 16$$
$$\log Q_{1v} = A_{1v} + B_{1v}T_e + C_{1v}T_e^2 + D_{1v}T_e^3 + F_{1v}T_e^4 - 16 \tag{B.10}$$

The coefficients A-F in the above expressions have been given in Table B.2, B.3 and B.4

Table B.2: Coefficients for calculating $Q_{0v}$ for $1500 \leq T_e \leq 6000$ K

| v | $A_{0v}$ | $B_{0v}$ K$^{-1}$ | $C_{0v}$ K$^{-2}$ | $D_{0v}$ K$^{-3}$ | $F_{0v}$ K$^{-4}$ | $\delta_{0v}$ |
|---|---|---|---|---|---|---|
| 1 | 2.025 | 8.782 $\times 10^{-4}$ | 2.954 $\times 10^{-7}$ | -9.562 $\times 10^{-11}$ | 7.252 $\times 10^{-15}$ | 0.06 |
| 2 | -7.066 | 1.001 $\times 10^{-2}$ | -3.066 $\times 10^{-6}$ | 4.436 $\times 10^{-10}$ | -2.449 $\times 10^{-14}$ | 0.08 |
| 3 | -8.211 | 1.092 $\times 10^{-2}$ | -3.369 $\times 10^{-6}$ | 4.891 $\times 10^{-10}$ | -2.706 $\times 10^{-14}$ | 0.10 |
| 4 | -9.713 | 1.204 $\times 10^{-2}$ | -3.732 $\times 10^{-6}$ | 5.431 $\times 10^{-10}$ | -3.008 $\times 10^{-14}$ | 0.10 |
| 5 | -10.353 | 1.243 $\times 10^{-2}$ | -3.850 $\times 10^{-6}$ | 5.600 $\times 10^{-10}$ | -3.100 $\times 10^{-14}$ | 0.13 |
| 6 | -10.819 | 1.244 $\times 10^{-2}$ | -3.771 $\times 10^{-6}$ | 5.385 $\times 10^{-10}$ | -2.936 $\times 10^{-14}$ | 0.15 |
| 7 | -10.183 | 1.185 $\times 10^{-2}$ | -3.570 $\times 10^{-6}$ | 5.086 $\times 10^{-10}$ | -2.769 $\times 10^{-14}$ | 0.15 |
| 8 | -12.698 | 1.309 $\times 10^{-2}$ | -3.952 $\times 10^{-6}$ | 5.636 $\times 10^{-10}$ | -3.071 $\times 10^{-14}$ | 0.15 |
| 9 | -14.710 | 1.409 $\times 10^{-2}$ | -4.249 $\times 10^{-6}$ | 6.058 $\times 10^{-10}$ | -3.300 $\times 10^{-14}$ | 0.15 |
| 10 | -17.538 | 1.600 $\times 10^{-2}$ | -4.916 $\times 10^{-6}$ | 7.128 $\times 10^{-10}$ | -3.941 $\times 10^{-14}$ | 0.15 |

## B.2.5 $O_2$ vibration

$$L_e(O_2) = n_e \, n(N_2) \, Q(T_e)[1 - \exp[2239(T_e^{-1} - T_n^{-1})]] \tag{B.11}$$

Table B.3: Coefficients for calculating Q$_{0v}$ for $300 \leq T_e \leq 1500$ K

| v | $A_{0v}$ | $B_{0v}$ K$^{-1}$ | $C_{0v}$ K$^{-2}$ | $D_{0v}$ K$^{-3}$ | $F_{0v}$ K$^{-4}$ | $\delta_{0v}$ |
|---|---|---|---|---|---|---|
| 1 | -6.462 | 3.151 $\times 10^{-2}$ | -4.075 $\times 10^{-5}$ | 2.439 $\times 10^{-8}$ | -5.479 $\times 10^{-12}$ | 0.14 |

Table B.4: Coefficients for calculating Q$_{1v}$ for $1500 \leq T_e \leq 6000$ K

| v | $A_{0v}$ | $B_{0v}$ K$^{-1}$ | $C_{0v}$ K$^{-2}$ | $D_{0v}$ K$^{-3}$ | $F_{0v}$ K$^{-4}$ | $\delta_{0v}$ |
|---|---|---|---|---|---|---|
| 2 | -3.413 | 7.326 $\times 10^{-3}$ | -2.200 $\times 10^{-6}$ | 3.128 $\times 10^{-10}$ | -1.702 $\times 10^{-14}$ | 0.11 |
| 3 | -4.160 | 7.803 $\times 10^{-3}$ | -2.352 $\times 10^{-6}$ | 3.352 $\times 10^{-10}$ | -1.828 $\times 10^{-14}$ | 0.11 |
| 4 | -5.193 | 8.360 $\times 10^{-3}$ | -2.526 $\times 10^{-6}$ | 3.606 $\times 10^{-10}$ | -1.968 $\times 10^{-14}$ | 0.12 |
| 5 | -5.939 | 8.807 $\times 10^{-3}$ | -2.669 $\times 10^{-6}$ | 3.806 $\times 10^{-10}$ | -2.073 $\times 10^{-14}$ | 0.08 |
| 6 | -8.261 | 1.010 $\times 10^{-2}$ | -3.039 $\times 10^{-6}$ | 4.318 $\times 10^{-10}$ | -2.347 $\times 10^{-14}$ | 0.10 |
| 7 | -8.185 | 1.010 $\times 10^{-2}$ | -3.039 $\times 10^{-6}$ | 4.318 $\times 10^{-10}$ | -2.347 $\times 10^{-14}$ | 0.12 |
| 8 | -10.823 | 1.199 $\times 10^{-2}$ | -3.620 $\times 10^{-6}$ | 5.159 $\times 10^{-10}$ | -2.810 $\times 10^{-14}$ | 0.09 |
| 9 | -11.273 | 1.283 $\times 10^{-2}$ | -3.879 $\times 10^{-6}$ | 5.534 $\times 10^{-10}$ | -3.016 $\times 10^{-14}$ | 0.09 |

with

$$\begin{aligned}
\log_{10}[Q(T_e)] = &-19.9171 + 0.0267 \, T_e - 3.9960 \times 10^{-5} \, T_e^2 + 3.5187 \times 10^{-8} \, T_e^3 \\
&- 1.9228 \times 10^{-11} \, T_e^4 + 6.6865 \times 10^{-15} \, T_e^5 - 1.4791 \times 10^{-18} \, T_e^6 \\
&+ 2.0127 \times 10^{-22} \, T_e^7 - 1.5346 \times 10^{-26} \, T_e^8 + 5.0148 \times 10^{-31} \, T_e^9
\end{aligned} \tag{B.12}$$

## B.2.6   O fine structure

$$\begin{aligned}
L_e(O) = n_e \, n(O) D^{-1} \Big( &S_{10}[1 - \exp[98.9(T_e^{-1} - T_n^{-1})]]\Big) + \\
&S_{20}\Big(1 - \exp[326.6(T_e^{-1} - T_n^{-1})]\Big) + \\
&S_{22}\Big(1 - \exp[227.7(T_e^{-1} - T_n^{-1})]\Big)
\end{aligned} \tag{B.13}$$

with

$$D = 5 + \exp(-326.6\ T_n^{-1}) + 3 \exp(-227.7\ T_n^{-1}),$$
$$S_{21} = 1.863 \times 10^{-11}$$
$$S_{20} = 1.191 \times 10^{-11}$$
$$S_{10} = 8.249 \times 10^{-16}\ T_e^{0.6} \exp(-227.7\ T_n^{-1})$$

(B.14)

## B.2.7   O($^1$D) excitation)

$$L_e(O(^1D)) = 1.57 \times 10^{-12}\ n_e\ n(O) \exp\left(d\frac{T_e - 3000}{3000\,T_e}\right)\left[exp\left(-22713\frac{T_e - T_n}{T_e\,T_n}\right) - 1\right]$$
$$d = 2.4 \times 10^4 + 0.3(T_e - 1500) - 1.947 \times 10^{-5}(T_e - 1500)(T_e - 4000)$$

(B.15)

## B.3   Ion cooling rates

Ions are cooled primarily by collisional energy exchange with neutral species, the rates for which are given in Table B.5.

Table B.5: Energy loss rates for ions due to collisions with neutrals

| Ion Mixture | Energy Loss rate ($10^{-14}$ eV cm$^{-3}$ s$^{-1}$) |
|:-----------:|:--------------------------------------------------:|
| $O^+$-$N_2$ | 6.6 n($O^+$) n($N_2$) ($T_i$ - $T_n$) |
| $O^+$-$O_2$ | 5.8 n($O^+$) n($O_2$) ($T_i$ - $T_n$) |
| $O^+$-$O$ | 0.21 n($O^+$) n($O$) ($T_i$ + $T_n$)$^{0.5}$ ($T_i$ - $T_n$) |
| $NO^+$-$N_2$ | 5.916 n($NO^+$) n($N_2$) ($T_i$ - $T_n$) |
| $NO^+$-$O_2$ | 5.45 n($NO^+$) n($O_2$) ($T_i$ - $T_n$) |
| $NO^+$-$O$ | 4.5n($NO^+$) n($O$) ($T_i$ - $T_n$) |
| $O_2^+$-$N_2$ | 5.807 n($O_2^+$) n($N_2$) ($T_i$ - $T_n$) |
| $O_2^+$-$O_2$ | 0.14 n($O_2^+$) n($O_2$) ($T_i$ + $T_n$)$^{0.5}$ ($T_i$ - $T_n$) |
| $O_2^+$-$O$ | 4.358 n($O_2^+$) n($O$) ($T_i$ - $T_n$) |

# B.4   Calculating Joule heating energy input

## B.4.1   Pedersen Conductivity

The expression for the Pedersen conductivity due to ions and electrons is given as:

$$\sigma_P = \sum_i \sigma_i \frac{\nu_i}{\nu_i^2 + \omega_i^2} + \sigma_e \frac{\nu_e}{\nu_e^2 + \omega_e^2} \tag{B.16}$$

where $\nu_i$, $\omega_i$, $\nu_e$ and $\omega_e$ are the ion and electron collision and cyclotron frequencies respectively, defined as

$$\nu_i = \sum_n \nu_{in}, \qquad \nu_{in} = C_{in} \, n_n \tag{B.17}$$

$$\nu_e = \sum_n \nu_{en} \tag{B.18}$$

$$\omega_i = \frac{q_i \, B}{m_i}, \qquad \omega_e = \frac{q_e \, B}{m_e} \tag{B.19}$$

$C_{in}$ are numerical coefficients that define non-resonant ion-neutral interactions. These have been given in Table B.6, while the expressions for resonant collision frequencies for are given in Table B.7. Table B.8 lists expressions for collision frequencies for electron - neutral interactions. Finally, the electron and ion conductivities are given as:

$$\sigma_e = \frac{n_e \, q_e^2}{m_e \, \nu_e}, \qquad \sigma_i = \frac{n_i \, q_i^2}{m_i \, \nu_i} \tag{B.20}$$

Table B.6: Collision frequencies ($C_{in} \times 10^{10}$) for nonresonant ion-neutral interactions

|         | N    | O    | N$_2$ | O$_2$ |
|---------|------|------|-------|-------|
| N$^+$   | -    | 4.42 | 7.47  | 7.25  |
| O$^+$   | 4.62 | -    | 6.82  | 6.64  |
| N$_2^+$ | 2.95 | 2.58 | -     | 4.49  |
| O$_2^+$ | 2.64 | 2.31 | 4.13  | -     |

Table B.7: Collision frequencies for resonant ion-neutral interactions. Densities are in cm$^{-3}$

| Species | $T_r$ (K) | $\nu^{-1}$ (s$^{-1}$) |
|---|---|---|
| N$^+$, N | $> 275$ | $3.83 \times 10^{-11} \; n(N) \; T_r^{0.5} \; (1 - 0.063 \log_{10} T_r)^2$ |
| O$^+$, O | $> 235$ | $3.67 \times 10^{-11} \; n(O) \; T_r^{0.5} \; (1 - 0.064 \log_{10} T_r)^2$ |
| N$_2^+$, N$_2$ | $> 170$ | $5.14 \times 10^{-11} \; n(N_2) \; T_r^{0.5} \; (1 - 0.069 \log_{10} T_r)^2$ |
| O$_2^+$, O$_2$ | $> 800$ | $2.59 \times 10^{-11} \; n(O_2) \; T_r^{0.5} \; (1 - 0.073 \log_{10} T_r)^2$ |

Table B.8: Collision frequencies for electron-neutral interactions. Densities are in cm$^{-3}$

| Species | $\nu^{-1}$ (s$^{-1}$) |
|---|---|
| N$_2$ | $2.33 \times 10^{-11} \; n(N_2) \; (1 - 1.21 \times 10^{-4} \; T_e) \; T_e$ |
| O$_2$ | $1.82 \times 10^{-10} \; n(O_2) \; (1 + 3.6 \times 10^{-2} \; T_e^{0.5}) \; T_e^{0.5}$ |
| O | $8.9 \times 10^{-11} \; n(O) \; (1 + 5.7 \times 10^{-4} T_e) T_e^{0.5}$ |

## B.5 Momentum transfer cross sections

The velocity averaged momentum cross sections that are used in section 2.8 to calculate electron thermal conductivity are given in units of cm$^{-2}$ below:

$$Q_{D,N_2}^- = 2.82 \times 10^{-17} \; T_e^{1/2} - 3.41 \times 10^{-21} T_e^{3/2} \tag{B.21}$$

$$Q_{D,O_2}^- = 2.2 \times 10^{-18} \; T_e^{1/2} + 7.92 \times 10^{-18} T_e^{1/2} \tag{B.22}$$

$$Q_{D,O}^- = 3.4 \times 10^{-16} \tag{B.23}$$

# B.6 Ambipolar diffusion coefficients

Expressions for the major and minor ion diffusion coefficients have been obtained from Schunk and Nagy (2009). For $O^+$, the diffusion parameters are given as:

$$D_a = \frac{2\ kT_p}{m_i\ \nu_i} \tag{B.24}$$

where

$$T_p = \frac{T_e + T_i}{2} \tag{B.25}$$

and $\nu_i$ is the ion collision frequency. Similarly, for $N^+$

$$D_a = \frac{kT_i}{m_i\ \nu_i} \tag{B.26}$$

The flux term in the continuity equation for ions undergoing diffusion includes a plasma scale height term, which is defined as

$$H_p = \frac{2\ kT_p}{m_i\ g} \tag{B.27}$$

# Appendix C

# Model Solar Fluxes

The ACE1D model follows the solar flux specification implemented in the TIE-GCM, after Solomon and Qian (2005). Given below are the wavelength ranges for each bin along with the corresponding values for the reference solar flux and the scaling parameter $A$. Details regarding the structure of the wavelength bins and calculation of the solar fluxes have been given Section 2.10.

Table C.1: **Solar Spectrum Parameters**

| Bin # | Wavelength Range (Å) | Reference Value (x$10^9$ ph cm$^{-2}$ s$^{-1}$) | A-factor |
|---|---|---|---|
| 1 | 0.5-4 | 5.010E-8 | 6.24E-1 |
| 2 | 4-8 | 1.000E-5 | 3.710E-1 |
| 3 | 8-18 | 2.000E-3 | 2.000E-1 |
| 4 | 18-32 | 2.850E-2 | 6.247E-2 |
| 5 | 32-70 | 5.326E-1 | 1.343E-2 |
| 6 | 70-155 | 1.270E+0 | 9.182E-3 |
| 7 | 155-224 | 5.612E+0 | 1.433E-2 |
| 8 | 224-290 | 4.342E+0 | 2.575E-2 |
| 9 | 290-320 | 8.380E+0 | 7.059E-3 |
| 10 | 320-540 | 2.861E+0 | 1.458E-2 |

C.1 – continued from previous page

| Bin # | Wavelength Range | Reference Value | A-factor |
|:---:|:---:|:---:|:---:|
| | (Å) | (x$10^9$ ph cm$^{-2}$ s$^{-1}$) | |
| 11 | 540-650 | 4.830E+0 | 5.857E-3 |
| 12 | 650-798 | 1.459E+0 | 5.719E-3 |
| 13 | 650-798 | 1.142E+0 | 3.680E-3 |
| 14 | 798-913 | 2.364E+0 | 5.310E-3 |
| 15 | 798-913 | 3.655E+0 | 5.261E-3 |
| 16 | 798-913 | 8.448E-1 | 5.437E-3 |
| 17 | 913-975 | 3.818E-1 | 4.915E-3 |
| 18 | 913-975 | 1.028E+0 | 4.955E-3 |
| 19 | 913-975 | 7.156E-1 | 4.422E-3 |
| 20 | 975-987 | 4.482E+0 | 3.950E-3 |
| 21 | 987-1027 | 4.419E+0 | 5.021E-3 |
| 22 | 1027-1050 | 4.235E+0 | 4.825E-3 |
| 23 | 1050-1100 | 3.298E+0 | 3.007E-3 |
| 24 | 1100-1150 | 3.200E+0 | 2.099E-3 |
| 25 | 1150-1200 | 8.399E+0 | 2.541E-3 |
| 26 | 1215.67 | 3.940E+2 | 4.230E-3 |
| 27 | 1200-1250 | 1.509E+1 | 3.739E-3 |
| 28 | 1250-1300 | 7.790E+0 | 2.610E-3 |
| 29 | 1300-1350 | 2.659E+1 | 2.877E-3 |
| 30 | 1350-1400 | 1.387E+1 | 2.632E-3 |
| 31 | 1400-1450 | 1.824E+1 | 1.873E-3 |
| 32 | 1450-1500 | 2.802E+1 | 1.202E-3 |
| 33 | 1500-1550 | 5.080E+1 | 1.531E-3 |
| 34 | 1550-1600 | 7.260E+1 | 1.125E-3 |

C.1 – continued from previous page

| Bin # | Wavelength Range | Reference Value | A-factor |
|:-----:|:----------------:|:---------------:|:--------:|
|       | (Å)              | (x10$^9$ ph cm$^{-2}$ s$^{-1}$) |          |
| 35    | 1600-1650        | 1.055E+2        | 1.043E-3 |
| 36    | 1650-1700        | 1.998E+2        | 6.089E-4 |
| 37    | 1700-1750        | 3.397E+2        | 5.937E-4 |

# Appendix D

# Numerical methods

The ACE1D model employs centered spatial derivatives and forward time derivatives to discretize and solve the continuity and energy equations discussed in Chapter 3. For the continuity equation, these take the form:

$$\frac{\partial \psi}{\partial Z} = \frac{\psi_{i+1}^{j+1} - \psi_{i-1}^{j+1}}{2\Delta Z} \tag{D.1}$$

$$\frac{\partial^2 \psi}{\partial Z^2} = \frac{\psi_{i+1}^{j+1} - 2\,\psi_{i+1}^{j} + \psi_{i+1}^{j+1}}{\Delta Z^2} \tag{D.2}$$

$$\frac{\partial \psi}{\partial t} = \frac{\psi_i^{j+1} - \psi_i^{j}}{\Delta t} \tag{D.3}$$

Where the derivatives are evaluated at the grid point $i$ and are used to solve for quantities at the timestep $j+1$. The equation then reduces to an expression in terms of the unknowns $\psi_{i-1}^{j+1}$, $\psi_i^{j+1}$ and $\psi_{i+1}^{j+1}$ with the general form:

$$p_i \psi_{i-1}^{j+1} + q_i \psi_i^{j+1} + r_i \psi_{i+1}^{j+1} = l_i \tag{D.4}$$

where $p_i$, $q_i$, $r_i$ and $l_i$ are known coefficients. Written for all grid points, this leads to the system of equations

$$
\begin{bmatrix}
q_1 & r_1 & 0 & 0 & \cdots & 0 \\
p_2 & q_2 & r_2 & 0 & \cdots & 0 \\
0 & p_3 & q_3 & r_3 & \cdots & 0 \\
\vdots & \vdots & \vdots & \vdots & \vdots & \vdots \\
0 & \cdots & 0 & 0 & p_n & q_n
\end{bmatrix}
\times
\begin{bmatrix}
\psi_1^{j+1} \\
\psi_2^{j+1} \\
\psi_3^{j+1} \\
\vdots \\
\psi_n^{j+1}
\end{bmatrix}
=
\begin{bmatrix}
l_1 \\
l_2 \\
l_3 \\
\vdots \\
l_n
\end{bmatrix}
\tag{D.5}
$$

which can then evaluated using tridiagonal matrix algorithms. Using boundary conditions and equations D.1 and D.2, grid points that fall outside of the model boundaries (e.g, $\psi_0^{j+1}, \psi_{n+1}^{j+1}$) are expressed in terms of quantities that are solved for in equation D.4.

## D.1 Block Tri-diagonal solver

In solving for the mass mixing ratios of the major species (Section 2.2), we solve for the coupled vector $\boldsymbol{\Psi} = [\psi_{O_2}, \psi_O]^T$, in which case the coefficients and variables in D.4 are matrices of dimensions $2 \times 2$ and $2 \times 1$ respectively. To solve for the vector $\boldsymbol{\Psi}$, we use the block tridiagonal algorithm:

$$
\begin{aligned}
&\textit{Forward elimination}: \\
&\boldsymbol{H_1} = \boldsymbol{Q_1^{-1} R_1} \\
&\boldsymbol{H_i} = -\left[\boldsymbol{Q_i + P_i H_{i-1}}\right]^{-1} \boldsymbol{R_i} \qquad (i = 2, 3, ..n) \\
&\boldsymbol{g_1} = \boldsymbol{Q_1^{-1} L_1} \\
&\boldsymbol{g_i} = \left[\boldsymbol{Q_i + P_i H_{i-1}}\right]^{-1} (\boldsymbol{L_i - P_i g_{i-1}}) \qquad (i = 2, 3, ..n)
\end{aligned}
\tag{D.6}
$$

$$
\begin{aligned}
&\textit{Backward substitution}: \\
&\boldsymbol{\Psi_n} = \boldsymbol{g_n} \\
&\boldsymbol{\Psi_i} = \boldsymbol{g_i + H_i \Psi_{i+1}} \qquad (i = n-1, ...2, 1)
\end{aligned}
\tag{D.7}
$$

where the unknown variables and the coefficients of D.5 are replaced by vectors and matrices denoted by $\boldsymbol{\Psi_i}, \boldsymbol{P_i}, \boldsymbol{Q_i}, \boldsymbol{R_i}$, and $\boldsymbol{L_i}$.

## D.2 Neutral gas cooling rates

The energy loss from the thermosphere due to radiative cooling by NO, $CO_2$ and $O(^3P)$ are temperature dependent, and results in the neutral gas heat equation (Section 2.6) being non-linear. Solving this equation thus requires linearization of these terms, which are of the general form:

$$
L = a \ e^{-b/T}
\tag{D.8}
$$

Where $T$ refers to the neutral temperature. Applying a Taylor series expansion to the above expression about the temperature at the current time step and ignoring the second order and higher terms, we obtain:

$$L = a \ e^{-b/T_n} + a \ e^{-b/T_n} \Big( \frac{-b}{T_n^2} \Big) (T_{n+1} - T_n) \tag{D.9}$$

where $T_n$ and $T_{n+1}$ refer to the neutral temperature at the current and next time step respectively. Grouping terms:

$$L = \Big[ a \ e^{-b/T_n} \Big( 1 - \frac{b}{T_n} \Big) \Big] + \Big[ - \frac{a \ b}{T_n^2} e^{-b/T_n} \ T_{n+1} \Big] \tag{D.10}$$
$$= L_1 + L_2$$

Here, $L_1$ comprises of known values, while $L_2$ is in terms of the unknown quantity $T_{n+1}$ that we wish to solve for. As a result, in the energy equation equivalent of D.4, $L_1$ is absorbed into the coefficient $l_i$ and can be referred to as the 'explicit' cooling term, while $L_2$ is a part of the coefficient $q_i$ and is referred to as an 'implicit' cooling term.

# Bibliography

Anicich, V. G. (2003). An index of the literature for bimolecular gas phase cation-molecule reaction kinetics.

Barth, C. A. (1992). Nitric oxide in the lower thermosphere. *Planetary and space science*, 40(2-3):315–336.

Baulch, D., Bowman, C. T., Cobos, C., Cox, R., Just, T., Kerr, J., Pilling, M., Stocker, D., Troe, J., Tsang, W., et al. (2005). Evaluated kinetic data for combustion modeling: supplement ii. *Journal of physical and chemical reference data*, 34(3):757–1397.

Berrington, K. and Burke, P. (1981). Effective collision strengths for forbidden transitions in en and eo scattering. *Planetary and Space Science*, 29(3):377–381.

Dilecce, G. and De Benedictis, S. (1999). Experimental studies on elementary kinetics in n2-o2 pulsed discharges. *Plasma Sources Science and Technology*, 8(2):266.

Dotan, I., Hierl, P. M., Morris, R. A., and Viggiano, A. (1997). Rate constants for the reactions of n+ and n2+ with o2 as a function of temperature (300–1800 k). *International journal of mass spectrometry and ion processes*, 167:223–230.

Dotan, I. and Viggiano, A. (1999). Rate constants for the reaction of o+ with no as a function of temperature (300–1400 k). *The Journal of chemical physics*, 110(10):4730–4733.

Duff, J., Dothe, H., and Sharma, R. (2003). On the rate coefficient of the n (2d)+ o2→ no+ o reaction in the terrestrial thermosphere. *Geophysical research letters*, 30(5).

Fell, C., Steinfeld, J., and Miller, S. (1990). Quenching of n (2 d) by o (3 p). *The Journal of Chemical Physics*, 92(8):4768–4777.

Fennelly, J., Torr, D., Richards, P., and Torr, M. (1994). Simultaneous retrieval of the solar euv flux and neutral thermospheric o, o2, n2, and temperature from twilight airglow. *Journal of Geophysical Research: Space Physics*, 99(A4):6483–6490.

Fox, J. L. and Sung, K. (2001). Solar activity variations of the venus thermosphere/ionosphere. *Journal of Geophysical Research: Space Physics*, 106(A10):21305–21335.

Hellberg, F., Rosén, S., Thomas, R., Neau, A., Larsson, M., Petrignani, A., and van der Zande, W. J. (2003). Dissociative recombination of no+: Dynamics of the x 1 $\sigma$+ and a 3 $\sigma$+ electronic states. *The Journal of chemical physics*, 118(14):6250–6259.

Herron, J. T. (1999). Evaluated chemical kinetics data for reactions of n (2 d), n (2 p), and n 2 (a 3 $\sigma$ u+) in the gas phase. *Journal of Physical and Chemical Reference Data*, 28(5):1453–1483.

Hierl, P. M., Dotan, I., Seeley, J. V., Van Doren, J. M., Morris, R. A., and Viggiano, A. (1997). Rate constants for the reactions of o+ with n2 and o2 as a function of temperature (300–1800 k). *The Journal of chemical physics*, 106(9):3540–3544.

Johnsen, R. and Biondi, M. A. (1980). Laboratory measurements of the o+ ($^2$d)+ n2 and o+ ($^2$d)+ o2 reaction rate coefficients and their ionospheric implications. *Geophysical Research Letters*, 7(5):401–403.

Kennealy, J., Del Greco, F., Caledonia, G., and Green, B. (1978). Nitric oxide chemiexcitation occurring in the reaction between metastable nitrogen atoms and oxygen molecules. *The Journal of Chemical Physics*, 69(4):1574–1584.

Lee, L. C., Slanger, T., Black, G., and Sharpless, R. (1977). Quantum yields for the production of o (1 d) from photodissociation of o2 at 1160–1770 å. *The Journal of Chemical Physics*, 67(12):5602–5606.

Li, X., Huang, Y.-L., Flesch, G., and Ng, C. (1997). A state-selected study of the ion–molecule reactions o+(4 s, 2 d, 2 p)+ n 2. *The Journal of chemical physics*, 106(4):1373–1381.

Logan, J. A., Prather, M., Wofsy, S., and McElroy, M. (1978). Atmospheric chemistry: Response to human influence. *Philosophical Transactions of the Royal Society of London A: Mathematical, Physical and Engineering Sciences*, 290(1367):187–234.

Midey, A. J., Miller, T. M., and Viggiano, A. (2004). Reactions of n+, n 2+, and n 3+ with no from 300 to 1400 k. *The Journal of chemical physics*, 121(14):6822–6829.

Midey, A. J. and Viggiano, A. (1999). Rate constants for the reaction of o 2+ with no from 300 to 1400 k. *The Journal of chemical physics*, 110(22):10746–10748.

Midey, A. J., Viggiano, A., Zhang, P., Irle, S., and Morokuma, K. (2006). A study of the reaction of n+ with o2: Experimental quantification of no+ (a 3+) production (298- 500 k) and computational study of the overall reaction pathways. *The Journal of Physical Chemistry A*, 110(9):3080–3086.

Peterson, J., Le Padellec, A., Danared, H., Dunn, G., Larsson, M., Larson, A., Peverall, R., Strömholm, C., Rosén, S., Af Ugglas, M., et al. (1998). Dissociative recombination and excitation of n 2+: Cross sections and product branching ratios. *The Journal of chemical physics*, 108(5):1978–1988.

Rawlins, W., Fraser, M., and Miller, S. (1989). Rovibrational excitation of nitric oxide in the reaction of oxygen with metastable atomic nitrogen. *The Journal of Physical Chemistry*, 93(3):1097–1107.

Richards, P. and Voglozin, D. (2011). Reexamination of ionospheric photochemistry. *Journal of Geophysical Research: Space Physics*, 116(A8).

Roble, R. G. (1995). Energetics of the mesosphere and thermosphere. *The upper mesosphere and lower thermosphere: a review of experiment and theory*, pages 1–21.

Sander, S., Golden, D., Kurylo, M., Moortgat, G., Wine, P., Ravishankara, A., Kolb, C., Molina, M., Finlayson-Pitts, B., Huie, R., et al. (2006). Chemical kinetics and photochemical data for use in atmospheric studies evaluation number 15. Technical report, Pasadena, CA: Jet Propulsion Laboratory, National Aeronautics and Space Administration, 2006.

Schunk, R. and Nagy, A. (2009). *Ionospheres: physics, plasma physics, and chemistry*. Cambridge university press.

Scott, G. B., Fairley, D. A., Freeman, C. G., McEwan, M. J., and Anicich, V. G. (1998). Gasphase reactions of some positive ions with atomic and molecular nitrogen. *The Journal of chemical physics*, 109(20):9010–9014.

Scott, G. B., Fairley, D. A., Milligan, D. B., Freeman, C. G., and McEwan, M. J. (1999). Gas phase reactions of some positive ions with atomic and molecular oxygen and nitric oxide at 300 k. *The Journal of Physical Chemistry A*, 103(37):7470–7473.

Sheehan, C. H. and St-Maurice, J.-P. (2004). Dissociative recombination of n2+, o2+, and no+: Rate coefficients for ground state and vibrationally excited ions. *Journal of Geophysical Research: Space Physics*, 109(A3).

Smithtro, C. and Solomon, S. (2008). An improved parameterization of thermal electron heating by photoelectrons, with application to an x17 flare. *Journal of Geophysical Research: Space Physics*, 113(A8).

Solomon, S. C. and Qian, L. (2005). Solar extreme-ultraviolet irradiance for general circulation models. *Journal of Geophysical Research: Space Physics*, 110(A10).

Stephan, A., Meier, R., Dymond, K., Budzien, S., and McCoy, R. (2003). Quenching rate coefficients for o+ (2p) derived from middle ultraviolet airglow. *Journal of Geophysical Research: Space Physics*, 108(A1).

Swaminathan, P., Strobel, D., Kupperman, D., Kumar, C. K., Acton, L., DeMajistre, R., Yee, J.-H., Paxton, L., Anderson, D., Strickland, D., et al. (1998). Nitric oxide abundance in the mesosphere/lower thermosphere region: Roles of solar soft x rays, suprathermal n (4 s) atoms, and vertical transport. *Journal of Geophysical Research: Space Physics*, 103(A6):11579–11594.

Watanabe, K. (1958). Ultraviolet absorption processes in the upper atmosphere. In *Advances in geophysics*, volume 5, pages 153–221. Elsevier.

Yonker, J. D. (2013). *Contribution of the first electronically excited state of molecular nitrogen to thermospheric nitric oxide*. PhD thesis, Virginia Polytechnic Institute and State University.

Zipf, E., Espy, P., and Boyle, C. (1980). The excitation and collisional deactivation of metastable n ($^2$p) atoms in auroras. *Journal of Geophysical Research: Space Physics*, 85(A2):687–694.